\documentclass[aps,prd,nofootinbib,superscriptaddress,preprintnumbers,longbibliography,eqsecnum,notitlepage,english]{revtex4-1}
\usepackage[export]{adjustbox}
\usepackage{blkarray}
\usepackage{amsfonts,amsmath,amssymb}
\usepackage{enumitem,mathtools}
\usepackage{array,multirow,booktabs,enumitem}
\usepackage{bbold,bbm,bm,nicefrac,dsfont}
\usepackage[dvipsnames,usenames]{color}
\usepackage{graphicx}
\usepackage{lipsum,float}
\usepackage[normalem]{ulem}
\usepackage[dvipsnames]{xcolor}
\usepackage[utf8]{inputenc}
\usepackage{soul}
\usepackage{hyperref}
\hypersetup{
    colorlinks = true,
    linkcolor = blue,
    urlcolor  = blue,
    citecolor = blue,
    anchorcolor = blue,
    pdftitle={The K(1460)...},
    pdfauthor={M Doering, K.P. Khemchandani, A. Martinez Torres}
}
\usepackage{cleveref} 
    \crefformat{figure}{Fig.~#2#1#3}
    \crefformat{table}{Tab.~#2#1#3}
    \crefformat{equation}{Eq.~(#2#1#3)}
    \crefformat{section}{Sect.~#2#1#3}

\newcommand{\ks}{$K(1460)$ }
\newcommand{\cg}[3]{(#1~#2|#3)}
\newcommand{\diff}{\mathrm{d}}
\newcommand{\non}{\nonumber \\}

\newcommand\TT{\rule{0pt}{2.6ex}}       
\newcommand\BBB{\rule[-1.6ex]{0pt}{0pt}} 
\graphicspath{{./plots/}}
\begin{document}
\title{Revisiting the three-kaon interaction and its relation with \texorpdfstring{$K(1460)$}{K(1460)}}

\author{M.~D\"oring}
\email{doring@gwu.edu}
\affiliation{Department of Physics, The George Washington University, Washington, DC 20052, USA}
\affiliation{Theory Center, Thomas Jefferson National Accelerator Facility, Newport News, VA 23606, USA}

\author{K. P. Khemchandani}
\email{kanchan.khemchandani@unifesp.br}
\affiliation{Universidade Federal de Sao Paulo, C.P. 01302-907, Sao Paulo, Brazil}

\author{A. Mart\'inez Torres}
\email{amartine@if.usp.br}
\affiliation{Universidade de Sao Paulo, Instituto de Fisica, C.P. 05389-970, Sao Paulo, Brazil.}
\preprint{JLAB-THY-25-4598}

\begin{abstract}
We test the hypothesis of \ks being a hadronic $J^P=0^-$ molecule through nonperturbative $S$-wave $K K\bar K$, $K\pi\pi$, $K\pi\eta$ coupled-channel dynamics in a three-body unitary isobar approach. The scalar two-body coupled-channel resonances $f_0(500)$, $f_0(980)$, $a_0(980)$ and $K^*_0(700)$ are generated in the subsytems in amplitudes that match phase shifts from experiment. In a first step, previous results in the limit of mass-degenerate, stable $f_0$ and $a_0$ isobars are reproduced. Once the full seven-coupled channel model is switched on, other $S$-matrix effects obscure and modify the resonance signal, including complex thresholds, a two-body cusp, and a triangle singularity at threshold.
\end{abstract}
\maketitle%

\section{Introduction}
\label{sec:introduction}
Since the discovery of the kaon in 1947~\cite{Rochester:1947mi}, understanding the kaon spectrum remains a challenge, with 24 states found in the mass range $\sim 700-3100$ MeV~\cite{ParticleDataGroup:2024cfk}, and with only approximately half of them considered established particles. While many of these states were discovered long ago~\cite{ParticleDataGroup:2024cfk}, there has been a renewed interest driven by modern experimental advances leading to signals of several new states, for example, at COMPASS~\cite{Wallner:2023rmn}, through diffractive production, or at LHCb~\cite{LHCb:2014kxv,LHCb:2017swu}, through weak decay processes of $B$ and $D$ mesons. Interestingly, even with the present facilities, despite having access to high-energy regions, no kaon resonances with masses higher than $3100$ MeV have been discovered although there are recent theoretical predictions supporting their existence~\cite{Ma:2018vhp,Ren:2018qhr,Ren:2018pcd,Di:2019jsx,Wu:2020job,Wei:2022jgc,Ozdem:2022ylm}. The approval of the European Organization for Nuclear Research (CERN) for the upgrade of the external M2 beam line to provide radio-frequency separated high-intensity and high-energy kaon and antiproton beams, in combination with a universal spectrometer of the COMPASS collaboration (Amber), is expected to drastically change the current status of kaon spectroscopy, mapping the entire spectrum with very high precision~\cite{Seitz:2023rqm}.

Among the kaon resonances listed by the Particle Data Group (PDG), there are only two $J^P=0^-$ states, the so-called $K(1460)$ and $K(1830)$, with the former being the only one considered to be established. In the case of $K(1460)$, since its first evidence in 1976 in a partial wave analysis of the $K\pi\pi$ invariant mass distribution of the $K^\pm p\to K^\pm \pi^+\pi^- p$ reaction~\cite{Brandenburg:1976pg}, few more experiments have supported its existence~\cite{ParticleDataGroup:2024cfk}. In Ref.~\cite{ACCMOR:1981yww} (a work which is more than 40 years old), the process $K^- p\to K^-\pi^-\pi^+ p$ was studied, and the corresponding partial-wave analysis of the $K\pi\pi$ invariant mass distribution suggested strong evidence for  the existence of a broad $J^P=0^-$ resonance at $\sim 1460$ MeV and a width of $\sim 260$ MeV. In Ref.~\cite{LHCb:2017swu}, the resonance structure in the process $D^0\to K^\mp \pi^\pm \pi^\pm\pi^\mp$ was investigated using $pp$ collision data with the LHCb experiment, finding contributions from the axial vector mesons $a_1(1260)$, $K_1(1270)$ and $K_1(1400)$ via $D^0\to a_1(1260)^+ K^-$ and $D^0\to K_1(1270/1400)^+\pi^-$ being the largest contributions to the decay amplitudes. A significant contribution was also found from the $\bar K(1460)$ state, confirming its resonant nature using a model-independent partial-wave analysis and finding partial decay fractions of the former state to $\bar K^*(892)\pi$ ($\sim 49\%-54\%$) and $(\pi\pi)_{S\text{-wave}}\bar K$ ($\sim 29\%-34\%$). The mass and width obtained for $K(1460)$ in Ref.~\cite{LHCb:2017swu} from the fit made to the data are $1482.40\pm3.58\pm15.22$ and $335.60\pm6.20\pm8.65$ MeV, respectively. In Ref.~\cite{Wallner:2023rmn}, the COMPASS collaboration studied strange pseudoscalar mesons via the $\rho(770) K$ decay mode in the process $K^- p\to K^-\pi^-\pi^+ p$. Their analysis shows preliminary results for a peak about $1400$ MeV in the $K\pi\pi$ invariant mass distribution which is described mainly by $K(1460)$. It is worth mentioning that the study of Ref.~\cite{Wallner:2023rmn} shows a signal at $\sim 1700$ MeV, $J^P=0^-$, as well, and which the authors related to the $K(1630)$ listed by the PDG. The quantum numbers of such a state are still not well established and its existence was claimed in the study of the $K^0_S\pi^+\pi^-$ invariant mass distribution performed (twenty-seven years ago) in Ref.~\cite{Karnaukhov:1998qq}. Recently, a $J^P=0^-$ tetraquark $ud\bar d\bar s$ nature has been claimed for this Kaon~\cite{Zhang:2025fuz}.

In the last fifteen years, there also seems to appear a boosted interest in describing the properties of $K(1460)$ through theoretical calculations long after the study of Ref.~\cite{Godfrey:1985xj}, in which, considering a relativistic quark model, \ks  was described as a $n^{2S+1}L_J=2^1 S_0$ excitation of the kaon. In recent times, the authors of Ref.~\cite{Pang:2017dlw}, modified the model of Ref.~\cite{Godfrey:1985xj} by taking the color screening effects into account and determined the mass spectrum of kaon resonances. To do this, the authors of Ref.~\cite{Pang:2017dlw} fixed the seven parameters of their model by selecting experimental information of some of the established kaon states listed by the PDG to optimize, by $\chi^2$ minimization, the masses of the considered kaon states. The $K(1460)$ state was one of the resonances introduced to fit the parameters in Ref.~\cite{Pang:2017dlw}, with the mass and quantum numbers obtained being 1457 MeV and $2^1 S_0$, respectively. Two-body partial decay widths for the \ks were also determined in Ref.~\cite{Pang:2017dlw}, finding a larger value (468 MeV) than the one listed by the PDG. In Ref.~\cite{Ebert:2009ub}, the masses of the ground state and the orbitally and radially excited states of quark-antiquark mesons made of $u$, $d$ and $s$ quarks were determined within a relativistic quark model based on the so-called quasipotential approach, in which the wave function of the bound quark-antiquark system satisfies a Schr\"odinger-type equation. The Regge trajectories for the angular and radial excitations were also obtained. A $2^1 S_0$ state with a mass of 1538 MeV was found and related to \ks. In Ref.~\cite{Taboada-Nieto:2022igy}, within a constituent quark model considering a one-gluon exchange force plus a quark-antiquark interaction based on the dynamical chiral symmetry breaking and color confinement, the masses of the kaon and of $K(1460)$ were determined, finding 481 and 1512 MeV, respectively, with the latter state being also interpreted as a $2^1 S_0$ excitation of the kaon. 

Some studies have suggested a different structure, a three-body molecular nature, for $K(1460)$, in which the interaction of $KK\bar K$ in the total isospin $1/2$ would be responsible for the generation of the mentioned state. Such a possibility sounds appealing, since the nominal mass for $K(1460)$ is close to the three-body $KK\bar K$ threshold ($\sim 1488$ MeV) and the $S$-wave interaction between the pairs is attractive in both isospins 0 and 1 for the case of the $K\bar K$ subsystems, generating the $f_0(980)$ and $a_0(980)$ resonances~\cite{vanBeveren:1986ea,Oller:1997ti,Oller:1998hw,Kaiser:1998fi,Hanhart:2007bd,Garcia-Martin:2011nna,Doring:2011vk,Bernard:2010fp}. To be more precise, in Ref.~\cite{MartinezTorres:2011gjk}, using two approaches, one based on the resolution of the Faddeev equations in momentum space, and other on the resolution of the Schr\"odinger equation considering Gaussian potentials to describe the $KK$ and $K\bar K$ interactions, the three-body $KK\bar K$ system was studied. A bump at $\sim 1420$ MeV with a width of $\sim 50$ MeV was observed in the modulus squared of the $T$-matrix for total isospin $1/2$ in the first case, and a state with mass of 1467 MeV and a width of 110 MeV was obtained in the second case. These structures were related to $K(1460)$. In Refs.~\cite{Filikhin:2020ksv,Filikhin:2023zjr}, the Faddeev equations in configuration space, using also a Gaussian form for the $KK$ and $K\bar K$ potentials, and the Coulomb interaction, were solved. A state with a mass $\sim 1460-1480$ MeV (depending on the values adopted for the parameters of the model) was found and associated with $K(1460)$. In Ref.~\cite{Zhang:2021hcl}, considering $f_0(980)$ and $a_0(980)$ to be $K\bar K$ bound states with the same mass and employing time ordered perturbation theory (TOPT), the $KK\bar K$ system in the isospin 1/2 and $J^P=0^-$ configuration was studied. Parameterizing the two-body subsystems as bound states with isobar fields coupling to $K\bar K$ describing  the $f_0$ and $a_0$ states, the $Kf_0(980)$ and $K a_0(980)$ coupled-channel scattering was studied by solving Lippmann-Schwinger type equations. A kernel in which the one-kaon exchange in the $t$- and $s$- channels moderates the interaction was considered to solve such equations. As a consequence, a bound state with a mass $\sim 2$ MeV below the $K f_0(980)/K a_0(980)$ nominal threshold was obtained. Other studies based on different methods have also predicted the \ks as a hadronic $KK\bar K$ molecule ~\cite{Albaladejo:2010tj,Kezerashvili:2014zza,Shinmura:2019nqw}.

The possible existence of a $K(1460)$ has also been studied within lattice QCD (see Refs.~\cite{Kronfeld:2012uk, Briceno:2017max} for reviews). In Ref.~\cite{Dudek:2010wm},  finite-volume energy levels were found in the vicinity of the \ks mass. Interestingly, this state shares its properties of an excited pseudoscalar state that decays into three particles with $\eta(1295)$ and $\pi(1300)$ -- even the masses of these states are similar taking into account the heavier strange quark. However, the  decay dynamics of \ks has not yet been explored on the lattice, in contrast to that of $\pi(1300)$, which was studied recently~\cite{Yan:2025mdm} through a finite-volume (FV), coupled-channel, three-body formalism referred to as ``Finite-Volume Unitarity'' (FVU)~\cite{Mai:2017bge}. See Refs.~\cite{Briceno:2025yuq, Dawid:2025doq, Garofalo:2022pux, Brett:2021wyd, Alexandru:2020xqf, Fischer:2020jzp, Hansen:2020otl, Guo:2020kph, Culver:2019vvu, Blanton:2019vdk, Mai:2019fba, Mai:2018djl} for other FV approaches containing analyses of lattice QCD data. Only few resonances decaying to three particles have been studied on the lattice yet, among them $a_1(1260)$~\cite{Mai:2021nul} and the ground state $\omega$~\cite{Yan:2024gwp}. We refer the reader to Refs.~\cite{Hansen:2019nir,Mai:2021lwb, Doring:2025sgb} for reviews and Refs.~\cite{Alotaibi:2025pxz, Hansen:2025oag, Draper:2023boj,Draper:2024qeh, Severt:2022jtg, Hansen:2020otl} for some recent finite-volume developments in the three-body sector including strangeness and unequal masses.

In this work, we study \ks in a framework related  to that of Ref.~\cite{Zhang:2021hcl}  where three kaon interactions are studied. Our approach has some different aspects: a) The isobar decay dynamics is taken into account through coupled channels in the isobar subsystem. In this work, the isobars are represented by two-body $t$-matrices obtained from the  coupled-channel Bethe-Salpeter equations with $S$-wave interactions in the so-called on-shell approximation~\cite{Oller:1997ti,Oller:1998hw}. Within this formalism, the low-lying scalar resonances [i.e., $f_0(500)$ (or $\sigma$), $f_0(980)$, $a_0(980)$, and $K^*_0(700)$ (or $\kappa$)] are generated from the lowest-order (LO) chiral dynamics~\cite{Oller:1997ti,Oller:1998hw}; b) To achieve a manifestly three-body unitary amplitude, three-body coupled-channels are required. Here, the coupled-channel system $KK\bar K$, $K\pi\pi$, and $K\pi\eta$ is studied through the spectator-isobar channels $K f^{K\bar K}_0$, $K f^{\pi\pi}_0$, $K a^{K\bar K}_0$, $K a^{\pi\eta}_0$, $\bar K K^{KK}_2$, $\pi\kappa^{K\pi}$, $\eta\kappa^{K\pi}$, where the superscript in the isobars represents their decay channels and the nomenclature $K_2$ is introduced to describe the (strangeness 2) $KK$ isobar. As one has new on-shell kinematics associated with light channels, the numerical aspects become more complex. On the other hand, we forego other aspects already studied in Ref.~\cite{Zhang:2021hcl}, like the influence of backward propagating particles.

The present approach is a ``dynamical coupled channel'' (DCC) model (see Ref.~\cite{Doring:2025sgb} for a review). This class of approaches is characterized by an off-shell scattering equation of the Lippman-Schwinger type with relativistic kinematics and the possibility to incorporate both two- and three-body dynamics~\cite{Mai:2017vot}, requiring dynamical particle exchange. The short-range dynamics can be modeled by hadronic Lagrangians or built in phenomenologically~\cite{Kamano:2011ih, Nakamura:2012xx}. In the light meson sector this approach has been recently applied to the $a_1(1260)$~\cite{Sadasivan:2020syi, Sadasivan:2021emk} and the $a_1(1420)$ triangle singularity~\cite{Sakthivasan:2024uwd}. See Refs.~\cite{Nakamura:2022rdd, Nakamura:2023obk, Nakamura:2023hbt, Zhang:2024dth, Shi:2024squ} for other unitary three-body coupled-channel applications. 
The DCC approach was also used to quantify three-body effects for exotic mesons  
like the $X(3872)$~\cite{Baru:2011rs}, the $Y(4260)$~\cite{Cleven:2013mka}, or the $T^+_{cc}$(3875)~\cite{Du:2021zzh,Lin:2022wmj} observed in the charmonium spectrum close to various three-body thresholds (see Ref.~\cite{Guo:2017jvc, Liu:2024uxn} for  related reviews).

\section{Formalism}
\label{sec:analytic-part}
The three-meson interaction is described as a $2+1$ process in which the quantum numbers of two interacting particles are combined in an ``isobar''. Then, a third ``spectator'' particle is added to the isobar to form the overall quantum numbers of the system. In the present case, we study the excited kaon spectrum with $J^P=0^-$ and isospin $I=1/2$ by considering only spinless isobars, and represent them by two-body scattering amplitudes. However, these amplitudes are not evaluated in their center-of-mass (c.m.) frame due to the presence of the third particle. 

The concept of coupled channels becomes useful because the strangeness $S=1$, $J^P=0^-$ state can be populated by different mesons. For a given set, like $K K \bar K$, one can still have different particle-isobar channels such as $Kf_0$ and $Ka_0$. Coupled-channel systems for isobars including spin are discussed in detail in Ref.~\cite{Feng:2024wyg} and are reviewed in Ref.~\cite{Doring:2025sgb}. 

An important aspect of the formalism concerns the implementation of manifest three-body unitarity  based on the conceptual work done in Ref.~\cite{Mai:2017vot}. In short, unitarity relates the imaginary parts of isobars to imaginary parts in the isobar-spectator interaction. Here, we rely on the unitary amplitude formulated in Ref.~\cite{Feng:2024wyg} with modifications required by the system under investigation as discussed in the next section.

\begin{figure}
\centering
\includegraphics[width=0.7\textwidth]{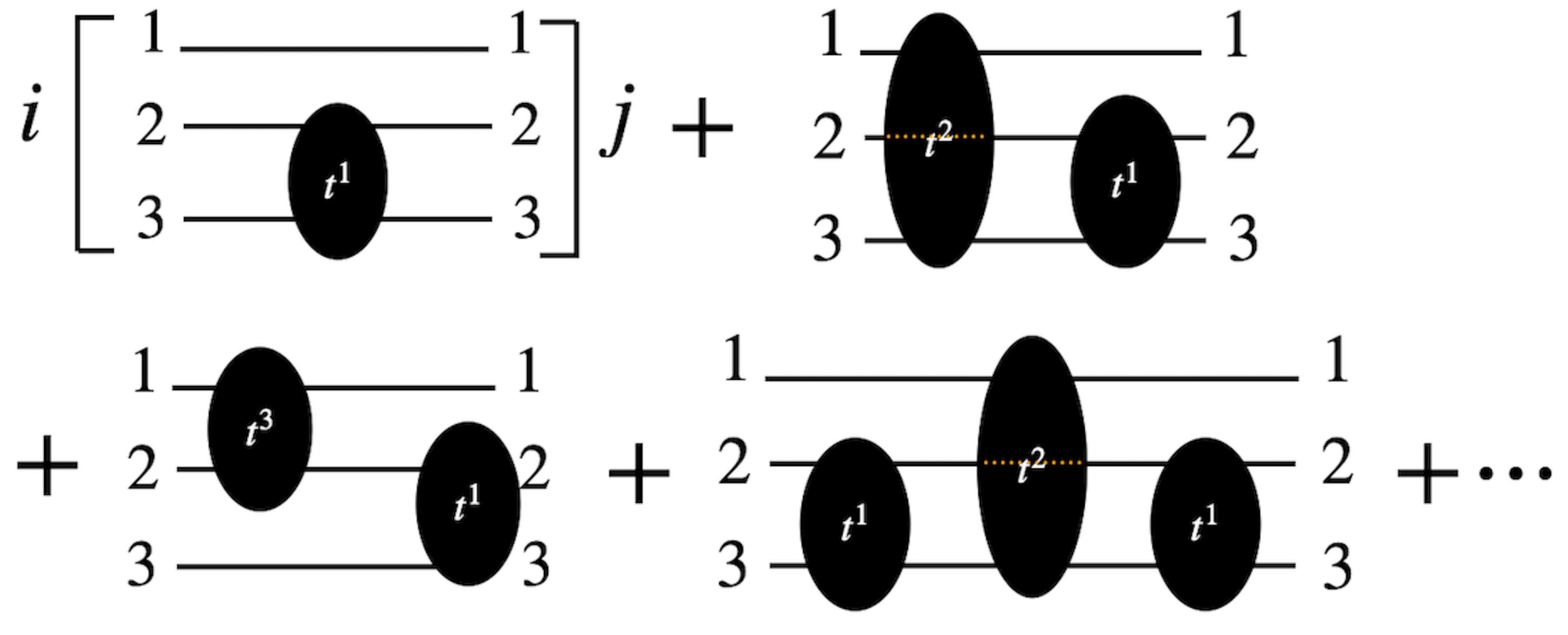}
\caption{Diagrammatic representation of some of the contributions considered in the $T^1$-partition. Particles are represented by horizontal lines. The blobs correspond to two-body interactions, described by two-body $t$-matrices, $t^i$ (with the superscript indicating the spectator particle). When writing the contribution associated with the diagrams, the latter are read from right to left, with the initial (final) state being represented by the ket $|i\rangle$ ($|j\rangle$). In this way, we get, in this case, the series $\langle j|(t^1+t^1 g^{12} t^2+t^1 g^{13} t^3+\cdots)|i\rangle$.}\label{FadDiagrams}
\end{figure}

It is worth mentioning at this point that the consideration of isobars as two-body $t$-matrices makes the analogy between the present formalism and that of solving the Faddeev equations for the considered three-body system quite straightforward. Before proceeding with further details of the present formalism, we find it interesting to dig into more details on this point. Within the Faddeev approach~\cite{Faddeev:1960su}, the three-body $T$-matrix for a system can be obtained by summing three partitions, $T^i$, $i=1,2,3$, satisfying 
\begin{align}
T^i=t^i+t^i g^{ij}T^j+t^i g^{ik}T^k,\label{FaEq}    
\end{align}
with $j\neq k\neq i=1,2,3$, $t^i$ being the two-body $t$-matrix that describes the interaction between the two-body subsystem obtained by excluding the particle $i$ of the three-body system, and $g^{ij}$ is the three-body Green's function representing the propagation of the particles involved between the interactions described by $t^i$ and $t^j$. The partition $T^i$ considers the different contributions to the scattering in which particle $i$ acts as a spectator in the last interaction (see Fig.~\ref{FadDiagrams}). Equation (\ref{FaEq}) represents a set of three coupled integral equations whose solution provides the three-body $T$-matrix of the system. An isospin basis can be considered to identify three-body states obtained from the dynamics involved. Typically, such a basis is characterized by the isospin of the three-body system, $I$, and that of one of the two-body subsystems, $I_\text{sub}$, and transitions like $\langle I,I^\prime_\text{sub}|T|I,I_\text{sub}\rangle$ are obtained~\cite{MartinezTorres:2007sr,Khemchandani:2008rk}. 

Continuing with the mentioned analogy, it is significant to observe that iterating in Eq.~(\ref{FaEq}), we find, for example,
\begin{align}
T^{1}&=t^1+t^1 g^{12}t^2+t^1 g^{12}t^2 g^{21}t^1+t^1 g^{12}t^2 g^{23}t^3+\cdots\nonumber\\
&\quad+t^1 g^{13}t^3+t^1 g^{13}t^3g^{31}t^1+t^1 g^{13}t^3 g^{32}t^2+\cdots\label{T1}
\end{align}
and similar expressions can be obtained for $T^2$ and $T^3$. Defining $T^{i}=t^i+T^{i}_R$, where $T^{i}_R$ considers contributions from connected diagrams, the terms on the right-hand side of Eq.~(\ref{T1}) associated with $T^1_R$ can be grouped into three different series based on the last two-body $t$-matrices appearing on them: $t^1(\cdots)t^1$, $t^1(\cdots)t^2$, $t^1(\cdots)t^3$, such that
\begin{align}
T^1_R&=t^1(g^{12}t^2g^{21}+g^{13}t^3 g^{31}+\cdots)t^1\nonumber\\
&\quad+t^1 (g^{12}+g^{13}t^{3}g^{32}+g^{12}t^2 g^{21}t^1 g^{12}+\cdots)t^2\nonumber\\
&\quad +t^1 (g^{13}+g^{13}t^3 g^{31}t^1 g^{13}+\cdots)t^3,\label{T1R}
\end{align}
and analogous expressions are found for $T^2_R$ and $T^3_R$. Next, given a three-body channel, denoted as (123), we can have three different reconfigurations of the particles, depending on which two particles interact first (indicated between brackets in the following notation): 1(23), 2(31), 3(12). In this way, the transition $1(23)\to 1(23)$, for example, considers contributions to the scattering in which the first and last two-body $t$-matrices involved are $t^1$, i.e., the series $t^1(\cdots)t^1$ in the first line of Eq.~(\ref{T1R}); similarly, the transition $2(31)\to 1(23)$ $[3(12)\to 1(23)]$ represents the contributions $t^1(\cdots)t^2$ $[t^1(\cdots)t^3]$ of Eq.~(\ref{T1R}). If we define these reconfigurations as channels 1, 2, 3 and construct the following matrices in this coupled channel space:
\begin{align}
 t=\left(\begin{array}{c|ccc}&1\equiv 1(23)&2\equiv 2(31)&3\equiv 3(12)\\\hline1\equiv 1(23)& t^1&0&0\\2\equiv 2(31)&0&t^2&0\\3\equiv 3(12)&0&0& t^3\end{array}\right),   
\end{align}
\begin{align}
 g=\left(\begin{array}{c|ccc}&1\equiv 1(23)&2\equiv 2(31)&3\equiv 3(12)\\\hline1\equiv 1(23)& 0&g^{21}&g^{31}\\2\equiv 2(31)&g^{12}&0&g^{32}\\3\equiv 3(12)&g^{13}&g^{23}& 0\end{array}\right),   
\end{align}
the contribution $T^i_R$, $i=1,2,3$, can be determined by summing the elements of the column $i$ of the matrix $\tilde{t}$ obtained from
\begin{align}
\tilde{t}=tgt+tgtgt+\cdots =t[g+gtg+gtgtg+\cdots]t\equiv t T t.   
\end{align}
By working in the ``$i(jk)$'' channel space, the $T^i_R$ contribution can be calculated by solving a Lippmann-Schwinger type equation (LSE) for the (amputated) $\tilde{T}$-matrix (more details are provided in the next sections), 
\begin{align}
T=g+gt T.\label{Ttilde}
\end{align}

\begin{figure}
\centering
\includegraphics[width=0.6\textwidth]{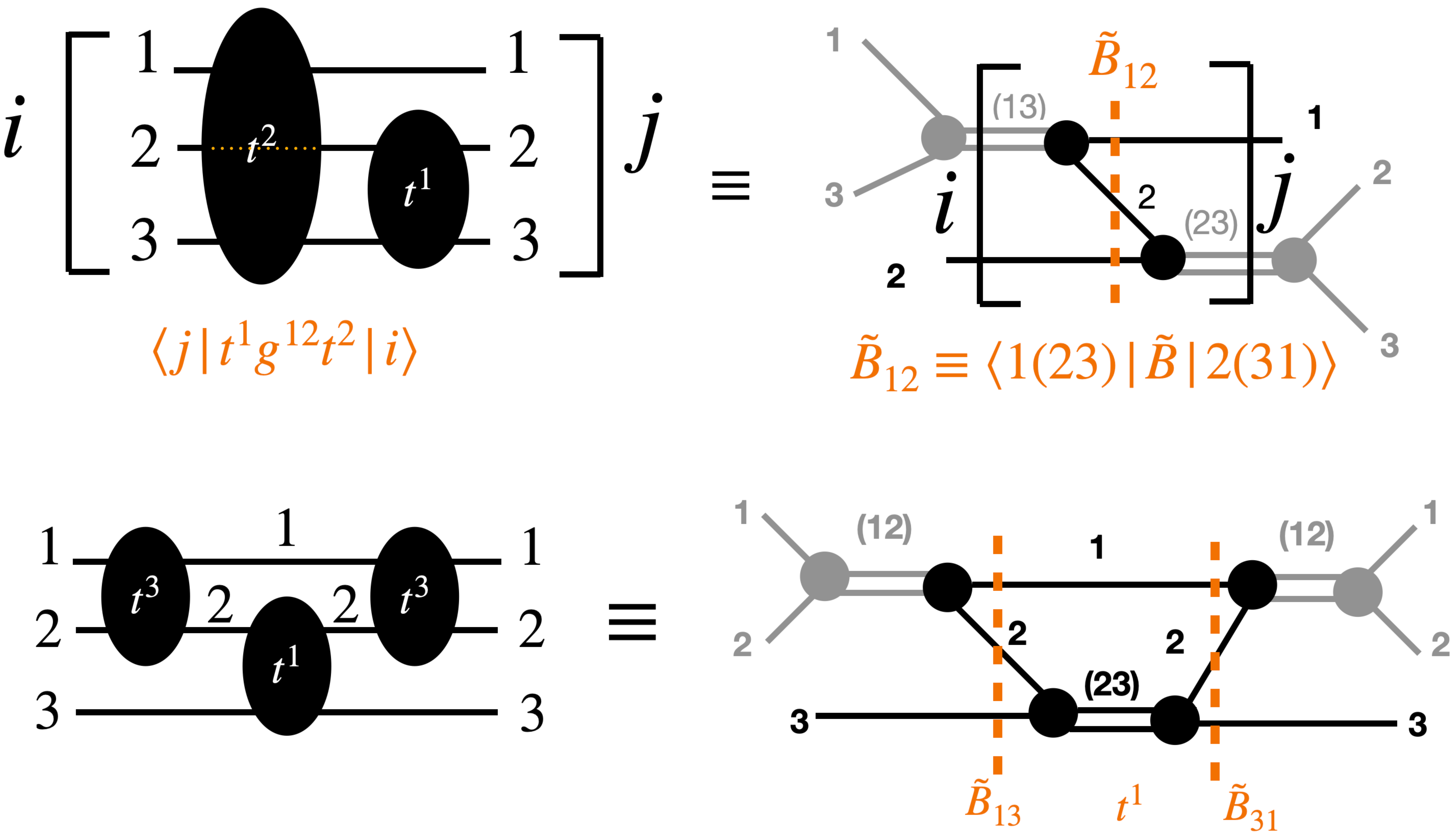}
\caption{Top panel, left: Diagram related to the contribution $t^1g^{12}t^2$ in the Faddeev approach. Top panel, right: The same diagram in terms of isobars (double line) to describe the interaction between particles 1 and 3 in the initial state, and 2 and 3 in the final state. Bottom row: Similar to the upper diagrams but for a case with one more interaction between the particles. The elements not included in the LSE, Eq.~(\ref{Ttilde}), namely the external isobars, are shown in gray.
}\label{equi}
\end{figure}

The association of two-body $t$-matrices with isobars implies, then, that each $i(jk)$ reconfiguration of a three-body channel is related to a spectator+isobar channel (see Fig.~\ref{equi}) and the contribution $\langle I,I^\prime_\text{sub}|T^i_R|I,I_\text{sub}\rangle$ is obtained by solving Eq.~(\ref{Ttilde}) in the isospin basis. To do this, isobars with a well-defined isospin are considered and combined with the spectator particle to find a certain isospin of the spectator-isobar system. This produces isospin factors which can be reabsorbed into the kernel of Eq.~(\ref{Ttilde}), changing it to $\tilde{B}$, as explained in the next sections. Also, decay channels for the isobars can be implemented in the formalism by replacing $t^i$ by $\tilde{\tau}_i$ (see Sec.~\ref{sip}).
Note that this is only possible for $S$-wave isobars. For isobars with higher partial waves there are  angular structures that are usually separated off the isobar  and included in the 3-body propagation ($\tilde B$ defined in the next section) which is then subject to a partial-wave projection~\cite{Sadasivan:2020syi, Feng:2024wyg}.
\subsection{Channel space and interactions}
In the present approach, the considered $S$-wave isobars are two-body subsystems in the form of full $2\to 2$ $t$-matrices in coupled channels. The isobars follow a positive strangeness-first convention. For example, the $f_0$ is constructed as a $K\bar K$ system and the $\kappa$ (short for the $K_0^*(700)$ quantum numbers) as a $K\pi$ system. Second, we adopt the convention of having the spectator before the isobar, i.e., $\pi\kappa$ and not $\kappa\pi$. Another typical convention, not followed here, is to have most positive strangeness to the left. For total isospin $I=1/2$, this convention would induce a minus sign compared to the current convention for the channels $\pi\kappa$ and $\bar K K_2$, where $K_2$ is the isospin $I_I=1$, strangeness $S_I=2$ $KK$ cluster (note that this interaction is repulsive).

We take into account two decay modes of $f_0$ and $a_0$, $f_0\to K\bar K,\pi\pi$ and $a_0\to K\bar K,\,\pi\eta$, while we assume that the $\kappa$ channel is elastic, $\kappa\to K \pi$, and omit its $K\eta$ mode. In other words, we allow for two-particle interactions $\pi\pi,\,\pi\eta,\,K\pi, K\bar K,\,KK$ but assume that the $K \eta$ and $\eta\eta$ interactions are negligible. Also, we neglect the small repulsive isospin $3/2$ $K \pi$ isobar. Note that the isospin $2$ $\pi\pi$ isobar cannot contribute to overall $I=1/2$. As a result, we include the channels $Kf_0,\,Ka_0,\,\bar KK_2,\,\pi\kappa$, and $\eta\kappa$.

In the present approach the re-arrangement process, sometimes referred to as $B$-term~\cite{Mai:2017vot} or $Z$-diagram~\cite{Matsuyama:2006rp, Doring:2025sgb}, is a particle $u$-channel exchange required by three-body unitarity~\cite{Mai:2017vot}: An incoming isobar decays. One of its decay products fuses with the incoming spectator to form the outgoing isobar. The other decay product becomes the outgoing spectator (see Fig.~\ref{equi}). This process allows all three particles to go on-shell in the right kinematic configuration, generating three-body singularities. In the partial-wave basis, these singularities lead to logarithmic branch points and associated cuts, see, e.g., a detailed discussion in Ref.~\cite{Feng:2024wyg}. The $B$ term in incoming (outgoing) channel $i(j)$ for incoming (outgoing) spectator momenta $\bm{p}(\bm{p}')$ reads~\cite{Feng:2024wyg}
\begin{align} 
    \tilde B_{ji}(s,\bm{p}',\bm{p})
    &=\frac{(\tilde {\bm I}_F)_{ji}\,
    v_j^*(p,P_3-p-p')
    v_i(p',P_3-p-p')}
    {2E_{p'+p}}\nonumber \\
    &\quad\times\left(\frac{1}{\sqrt{s}-E_{p}-E_{p'}-E_{p'+p})+i\epsilon}+\frac{1}{\sqrt{s}-{\cal E}_{p}-{\cal E}_{p'}-E_{p'+p}}\right)
    \label{btilde}
\end{align}
with $v_{i}(p,q)=1$ for $S$-wave isobars, and $P_3$, $p$, $p^\prime$ being the total four-momentum of the system and that of the spectator particle in the initial and final states, respectively. Here, $E_q$ is the (channel dependent) energy of the spectator or exchanged particle at momentum $\bm q$ while ${\cal E}$'s indicate the energies of the isobars that appear in the diagram with backward propagating exchange particle, added by using the rules of TOPT. This term is generally very small, as tested. 
We evaluate the amplitude in the three-body c.m. frame, $P_3=(\sqrt{s},\bm{0})$. In this work we continue using a ``tilde'' notation introduced in Ref.~\cite{Feng:2024wyg} to distinguish quantities from the ones used in earlier literature, the main difference concerning the location of symmetry and isospin factors in either interaction ($B$) or isobar-spectator propagator ($\tau$, defined below).

We study here the possibility of obtaining kaon resonances as an emergent phenomena from the long-range meson exchange process and its non-linear resummation in a scattering equation. Therefore, we neglect other interactions, called $C$-term in Ref.~\cite{Mai:2017vot}, that respect unitarity as long as they are real-valued in the physical region. Explicit $s$-channel~\cite{Sadasivan:2021emk} and $t$-channel exchanges and/or effective contact interactions can provide a microscopic picture for these short-range interactions. Once data are fitted, the $C$-term becomes important, see, e.g., Refs.~\cite{Brett:2021wyd, Sadasivan:2020syi, Mai:2021nul}.
In dynamical coupled-channel approaches, the interaction is often modeled by a substantial set of $s$, $t$ and $u$-channel diagrams arising from Lagrangians, especially in the meson-baryon sector~\cite{Matsuyama:2006rp, Ronchen:2012eg, Ronchen:2014cna, Kamano:2019gtm} as reviewed recently in Ref.~\cite{Doring:2025sgb} (see also Ref.~\cite{Doring:2025wms} for an introduction).

We also neglect the kaon $s$-channel exchange. The kaon appears as a bound state in the $J^P=0^-$ channel and undergoes renormalization as carried out in Ref.~\cite{Zhang:2021hcl}. The kaon is about a GeV lighter than the excited kaon region we are interested in which justifies this approximation. Note also that there are additional ingredients such as backwards propagating spectator mesons that also lead to small (but not tiny) corrections that were quantified in Ref.~\cite{Zhang:2021hcl}.

Isospin coefficients are calculated as
\begin{align}
\tilde I_F=\sum_{\substack{m,n,
m',n'\\
|s_x|=S_x} }&\big[
\cg{I_{S'}m'}{I_{I'}n'}{II_3}\,
c(I_{I'},I_x,I_S,S_S,S_{I'})\,
\cg{I_xn-m'}{I_{S}m}{I_{I'}n'} 
\non
&\times
\delta_{S_I+S_{I'},S+s_x}\,
c(I_{I},I_x,I_{S'},S_{S'},S_{I})\,
\cg{I_xn-m'}{I_{S'}m'}{I_In}\,
\cg{I_Sm}{I_In}{II_3}
\big]
\label{eq:isostrange}
\end{align}
with total strangeness chosen as $S=+1$, strangeness of the exchanged particle, $S_x\ge 0$, incoming isobar/spectator strangeness $S_I/S_S$, outgoing isobar/spectator strangeness $S_{I'}/S_{S'}$, total/incoming isobar/outgoing isobar/exchange-particle isospin $I/I_I/I_{I'}/I_x$, isospin third components $I_3,n,m,n',m'$, and
\begin{align}
c(I_{I'},I_x,I_{S},S_{S},S_{I'})=\begin{cases}
    (-1)^{I_{I'}-I_x-I_{S}}&\text{if } 2S_S>S_{I'} \ ,\\
    +1&\text{else.}
    \label{cstrange}
\end{cases}
\end{align}
The first line in Eq.~\eqref{eq:isostrange} contains the isospin Clebsch-Gordan (CG) factor coupling outgoing spectator and isobar to external, total isospin in the chosen spectator-first convention. The first line also contains an internal isospin-CG factor coupling the exchanged particle and the incoming spectator to the outgoing isobar. The factor $c$ corrects the order in that isobar-decay coefficient to ensure the first particle in it has the larger or equal strangeness. The order of the particles in the isobar amplitude $\tilde\tau$ (defined in Sec.~\ref{sec:isobars}) must correspond to this chosen order, i.e., $K\bar K$. The factor $c$ is negative for the $f_0$ isobar coupling to a $K^0$ or $K^+$ spectator, or for the $\kappa^+$ isobar coupling to a $K^0$ or $K^+$ spectator. Note that the condition $2S_S>S_{I'}$ in Eq.~\eqref{cstrange} arises from the condition $S_S>s_x$ and $s_x=S_{I'}-S_S$. 

The second line in Eq.~\eqref{eq:isostrange} contains a strangeness-conserving Kronecker delta restricting the sum over $s_x$, the internal isospin-CG that couples the exchanged  particle and the outgoing spectator to the incoming isobar, the pertinent correction factor to ensure the positive-strangeness first convention, and the external isospin-CG coupling the incoming spectator and isobar to the total isospin. All phases for involved particles cancel (usually chosen as negative for $\pi^+$, $K^-$). 
In summary, the isospin factors from Eq.~\eqref{eq:isostrange} to be used in Eq.~\eqref{btilde} are 
\begin{align}
   \tilde {\bm I}_F (I=1/2)=
   \ \ 
   \begin{blockarray}{ccccccccc}
   Kf_0^{K\bar K}&Kf_0^{\pi\pi}&Ka_0^{K\bar K}&Ka_0^{\pi\eta}&\bar KK_2^{KK}&\pi\kappa^{K\pi}&\eta\kappa^{K\pi}\\
       \begin{block}{(ccccccccc)}
        \nicefrac{1}{2}&0&\frac{\sqrt{3}}{2}&0&-\nicefrac{\sqrt{3}}{2}&0&0
        \\
        0&0&0&0&0&\nicefrac{1}{\sqrt{3}}&0
        \\
        \frac{\sqrt{3}}{2}&0&-\nicefrac{1}{2}&0&-\nicefrac{1}{2}&0&0
        \\
        0&0&0& 0&0&0&1
        \\
        -\nicefrac{\sqrt{3}}{2}&0&-\nicefrac{1}{2}&0&0&0&0 
        \\
        0&\nicefrac{1}{\sqrt{3}}&0&0&0& -\nicefrac{1}{3}&0
        \\   
        0&0&0&1&0&0& 0
        \\
    \end{block}
    \end{blockarray}\ \ .
    \label{isoexplicit}
\end{align}
The superscript in the isobars indicates their decay mode. 
Note that many coefficients vanish because the isobar decay modes in the transitions do not match. For example, there is kaon exchange for the $Kf_0\to Ka_0$ transition, but no pion exchange, i.e., the $(1,4)$ and $(4,1)$ elements vanish. 

\subsection{Scattering equation, isobars, and production amplitude}\label{sip}

The plane-wave equation from Ref.~\cite{Feng:2024wyg} for the isobar-spectator scattering $T$-matrix reads
 \begin{align}
    T_{ji}(s,{\bm p}',\bm{p})=
    \tilde B_{ji}(s,{\bm p}',\bm{p})+
    \int\frac{\mathrm{d}^3l}{(2\pi)^3\,2E_{l}}
    \tilde B_{jk}(s,{\bm p}',{\bm  l})
    \,
    \tilde \tau_{kk'}(\sigma_l) \, 
     T_{k'i}(s,{\bm l},{\bm{p}}) \ ,
    \label{eq:T3-integral-equation}
\end{align}
with spectator energy in the loop $E_l=\sqrt{m_k^2+l^2}$, spectator mass $m_k=m_{k'}$, the total Mandelstam $s$, and the isobar Mandelstam $\sigma$ sub-energy squared related by
\begin{align}
    s=P_3^2,\quad \sigma_l=(P_3-p)^2=s+m_k^2-2\sqrt{s}E_{l}.
    \label{eq:sigma}
\end{align}  
This scattering equation is of Lippman-Schwinger type with relativized kinematics. It iterates the $\tilde{B}$-term to all orders, with isobar spectator propagation $\tilde\tau$ specified in the next section. Channel and momenta indices correspond to the ones introduced in Eq.~\eqref{btilde}.
Note that Eq.~\eqref{eq:T3-integral-equation} is obtained from a (covariant) Bethe-Salpeter equation
by maintaining only the positive-energy component of the spectator and putting it on-shell, ${\tilde\tau(\sigma_l)=2\pi\delta(l^2-m_k^2)\theta(l^0)\,S}$, where $S$ is the generic isobar-spectator propagation~\cite{Mai:2017vot}. 

After projection of Eq.~\eqref{eq:T3-integral-equation} following the procedure and conventions defined in Ref.~\cite{Feng:2024wyg} (see also Refs.~\cite{Briceno:2024ehy, Jackura:2023qtp}), the isobar-spectator equation reads
\begin{align}
     T_{ji}(s,p',p)&
    =\tilde B_{ji}(s,p',p)+
    \int\limits_0^\Lambda 
    \frac{\text{d}l\,l^2}{(2\pi)^3\,2E_{l}}
    \tilde B_{jk}(s,p',l)
    \,
    \tilde \tau_{kk'}(\sigma_l) \,
     T_{k'i}(s,l,p) \ ,
    \label{eq:TLL}
\end{align}
where $i,j,k,k'$ are the channel indices in the JLS basis according to Eq.~\eqref{isoexplicit}.
The integration contour is referred to as spectator momentum contour (SMC) in the following.
This equation  is formulated in the product space of five isobar-spectator channels, two sub-channels for $f_0$ and $a_0$ each,  and  spectator momenta. With the channel ordering of Eq.~\eqref{isoexplicit}, the isobar propagator $\tilde\tau$ of Eqs.~\eqref{eq:T3-integral-equation} and \eqref{eq:TLL} acquires a block-diagonal structure, 
\label{sec:tau}
\begin{align}
    \tilde\tau=\text{diag}\left(
        \begin{pmatrix}
        \tilde\tau_{Kf^0}^{K\bar K\leftarrow K\bar K} &  \tilde\tau_{Kf_0}^{\pi\pi\leftarrow K\bar K} \\
        \tilde\tau_{Kf_0}^{ K\bar K\leftarrow\pi\pi} & 
        \tilde\tau_{Kf_0}^{ \pi\pi\leftarrow\pi\pi}
        \end{pmatrix},
        \begin{pmatrix}
        \tilde\tau_{Ka^0}^{K\bar K\leftarrow K\bar K} &  \tilde\tau_{Ka_0}^{\pi\eta\leftarrow K\bar K} \\
        \tilde\tau_{Ka_0}^{ K\bar K\leftarrow\pi\eta} & 
        \tilde\tau_{Ka_0}^{ \pi\eta\leftarrow\pi\eta}
        \end{pmatrix},
        \tilde\tau_{\bar KK_2}^{KK\leftarrow KK},
        \tilde\tau_{\pi\kappa}^{K\pi\leftarrow K\pi},
        \tilde\tau_{\eta\kappa}^{K\pi\leftarrow K\pi}
    \right) \ ,
\end{align}
in which each element is a diagonal matrix of complex spectator momenta.

\begin{figure}
\centering
\includegraphics[width=0.9\textwidth]{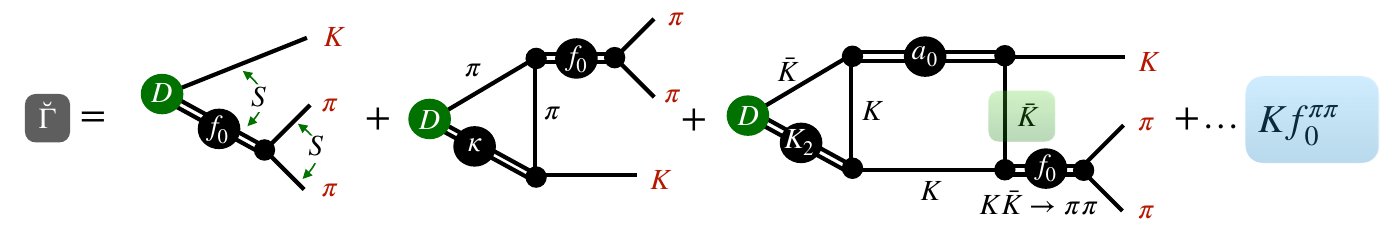}
\caption{A typical production process of $Kf_0^{\pi\pi}$ including the ``disconnected'' piece and the rescattering part. The $Kf_0$ state can be populated through different intermediate channels. The first diagram indicates that all three particles are in relative $S$-wave. The last diagram shows a near on-shell kaon exchange (highlighted in green) followed by a channel transition $K\bar K\to\pi\pi$.}
\label{fig:production}
\end{figure}

Most physical processes involving three stable particles in the final state occur as production reactions, for which the $T$-matrix provides the final-state interaction (FSI). Following Ref.~\cite{Feng:2024wyg}, we first define a quantity
\begin{align}
    \tilde\Gamma_j (s,p')=\int_{\Gamma}\frac{\diff p\,p^2}{(2\pi)^3\,2E_{p}}\, T_{ji}(s,p',p)\tilde\tau_{ik}(\sigma(p))D_k(s,p) \ ,
    \label{eq:normalgamma}
\end{align}
where $D_k(s,p)$ is a real-valued momentum and energy-dependent ``elementary'' production process that can be fit to data. It should obey minimal constraints like the correct centrifugal barrier~\cite{Feng:2024wyg}. In the present study that only considers $S$-wave isobar-spectator interactions, $D=1$ for all channels. The quantity $\tilde \Gamma$ corresponds to a production process $D$ followed by the FSI. For a full physical process one needs to add a final isobar and its decay vertex, as well as a ``disconnected'' piece that only contains an isobar propagation but no FSI. This defines~\cite{Feng:2024wyg}
\begin{align}
    \breve\Gamma_{j}(s,p')=\breve v_{j}(\sigma(p'))\,\tilde\tau_{ji}(\sigma(p'))\left[\tilde\Gamma_{i}(s,p')+D_{i}(s,p')\right]
    \label{eq:gammabrev} \ .
\end{align}
There is no angular structure in the $S$-wave isobar decays so that $\breve v_{j}=1$. 
In Fig.~\ref{fig:production} a typical production process $\breve\Gamma$ is shown, for the $\pi\pi K$ final state produced through $Kf_0$.
The $\pi\pi K$ final  state is of particular interest because the $K(1460)$ should be visible in it, in contrast to the heavier final state $KK\bar K$.

\subsection{Two-body input}
\label{sec:isobars}
To determine the two-body $t$-matrices ($\tilde{\tau}$) describing the interactions between the different pseudoscalar subsystems, we follow the approach of Refs.~\cite{Oller:1997ti,Oller:1998hw}, in which effective Lagrangians based on the chiral symmetry and its spontaneous breaking are considered. In particular, in Ref.~\cite{Oller:1997ti}, the lowest order chiral Lagrangian was used, while in Ref.~\cite{Oller:1998hw} next-to-leading order contributions were included, finding compatible results with those of Ref.~\cite{Oller:1997ti}. In this work, we use the lowest order chiral Lagrangian, as done in Ref.~\cite{Oller:1997ti}, and calculate the related amplitudes $V_{ji}$ (here $i$/$j$ refer to incoming/outgoing two-body channels), which are further projected onto the $S$-wave and into the isospin basis for those coupled channels relevant in the energy region to be studied (we refer the reader to the Appendix B of Ref.~\cite{Oller:1998hw} and that of Ref.~\cite{Malabarba:2023zez} to find the corresponding expressions for $v_{ji}$). 
Note that, in contrast to the normalization used there, we do not absorb symmetry factors ${\cal N}_i$ in the definition of states but in the loops, see below. In other words, our transition amplitudes are larger by $V_{ji}=1/\sqrt{{\cal N}_j{\cal N}_i}\,v_{ji}$ than the transitions $v_{ji}$ from the above references.
A matrix $V(\sigma)$ is then constructed in this coupled channel space and used as kernel to solve the Bethe-Salpeter equation in its on-shell factorized form~\cite{Oller:1997ti,Oller:1998hw}:
\begin{align}
\tilde{\tau}=V+VG\tilde{\tau}=[1-V(\sigma) G(\sigma)]^{-1} V(\sigma),\label{BSon}
\end{align}
where $G(\sigma)$ is a matrix whose elements are the loop functions for the possible intermediate states in the mentioned coupled-channel space:
\begin{align}
G_i(\sigma)=\mathcal{N}_i\int\limits_0^\Lambda  \frac{\diff q\,q^2}{(2\pi)^2}\frac{\omega_{1i}(q)+\omega_{2i}(q)}{\omega_{1i}(q)\omega_{2i}(q)[\sigma-(\omega_{1i}(q)+\omega_{2i}(q))^2+i\epsilon]}, \label{Gloop}   
\end{align}
with $\omega_{1(2)i}(q)=\sqrt{q^2+m^2_{1(2)i}}$ being the center-of-mass energies of the two propagating pseudoscalars, with respective masses $m_{1i}$ and $m_{2i}$, in the channel $i$. The normalization factor $\mathcal{N}_i$ is equal to 1 (1/2) when the two pseudoscalars constituting the channel $i$ are distinguishable (indistinguishable). Note that pions count as indistinguishable in the isospin basis, irrespective of their charge. Also, note that in Ref.~\cite{Feng:2024wyg} the chiral unitary isobars $t_{ji}$ were taken from the literature with the normalization ${\cal N}_i$ absorbed in the states, which required the correction $\tilde\tau_{ji}=t_{ji}/\sqrt{{\cal N}_j{\cal N}_i}$. 

The integral in Eq.~(\ref{Gloop}) diverges and needs to be regularized with either a cut-off $\Lambda$, as explicitly shown in Eq.~(\ref{Gloop}), or by using dimensional regularization~\cite{Oller:1998hw}, introducing in $G_i$ a subtraction constant $a_i$ at a certain energy scale $\mu$. In our case, we use the dimensional regularization scheme, with~\cite{Oller:1998hw}
\begin{align}
G_i&=\frac{1}{16\pi^2}\Bigg\{a_i(\mu)+\text{ln}\frac{m^2_{1i}}{\mu^2}+\frac{m^2_{2i}-m^2_{1i}+\sigma}{2\sigma}\text{ln}\frac{m^2_{2i}}{m^2_{1i}}+\frac{p_i}{\sqrt{\sigma}}\Bigg[\text{ln}(\sigma-m^2_{1i}+m^2_{2i}+2p_i\sqrt{\sigma})\nonumber\\
&\quad+\text{ln}(\sigma+m^2_{1i}-m^2_{2i}+2p_i\sqrt{\sigma})-\text{ln}(-\sigma+m^2_{1i}-m^2_{2i}+2p_i\sqrt{\sigma})-\text{ln}(-\sigma-m^2_{1i}+m^2_{2i}+2p_i\sqrt{\sigma})\Bigg]\Bigg\},\label{Gi}
\end{align}
where $p_i$ is the center-of-mass linear momentum for the two mesons in the channel $i$, and $a_i(\mu)\sim -1$ for $\mu\sim 1200$ MeV. These parameters are fixed to reproduce the observed two-body $S$-wave phase shifts and inelasticities in the isospin 0, 1 and $1/2$ sectors for the coupled channels considered, as done in Refs.~\cite{Oller:1997ti,Oller:1998hw}. 

As mentioned earlier, in the case of the $K\pi$ interaction in isospin $1/2$, we neglect the contributions to $K\pi$ scattering arising from the $K\eta$, and $K\eta^\prime$ inelastic channels.  Also, the $K\pi\to K\pi$ $V$-amplitude  projected on the $S$-wave diverges as $1/\sigma$ when $\sigma\to 0$. Although this is not a problem when using the two-body $t$-matrix to reproduce experimental data on $K\pi$ scattering, it becomes one when using this amplitude to solve Eq.~(\ref{eq:TLL}) (for which we need $\sigma<0$), since an artificial pole at $\sigma=0$ would appear in the complex energy plane. To avoid this problem, the term $1/\sigma$ present in the mentioned amplitude is replaced by its series expansion around the $K\pi$ threshold. With these changes, to reproduce the experimental isospin-$1/2$ $K\pi$ $S$-wave phase shift up to energies $\sim 1200$ MeV, we use different weak decay constants for the $\pi$ and $K$ mesons, with $f_\pi=93$ MeV and $f_K=1.22f_\pi$~\cite{Gasser:1984gg}, instead of considering $f_\pi=f_K$, as sometimes done. It should be mentioned  that, since we consider two-body $t$-matrices as isobars, the order in which particles are combined to get isospin kets should be consistent with the order followed to combine the particles coupled to the isobars (see Fig.~\ref{equi}) to determine the isospin coefficients present in Eq.~(\ref{isoexplicit}).

\begin{figure}[htb]
\begin{center}
\includegraphics[width=0.55\textwidth]{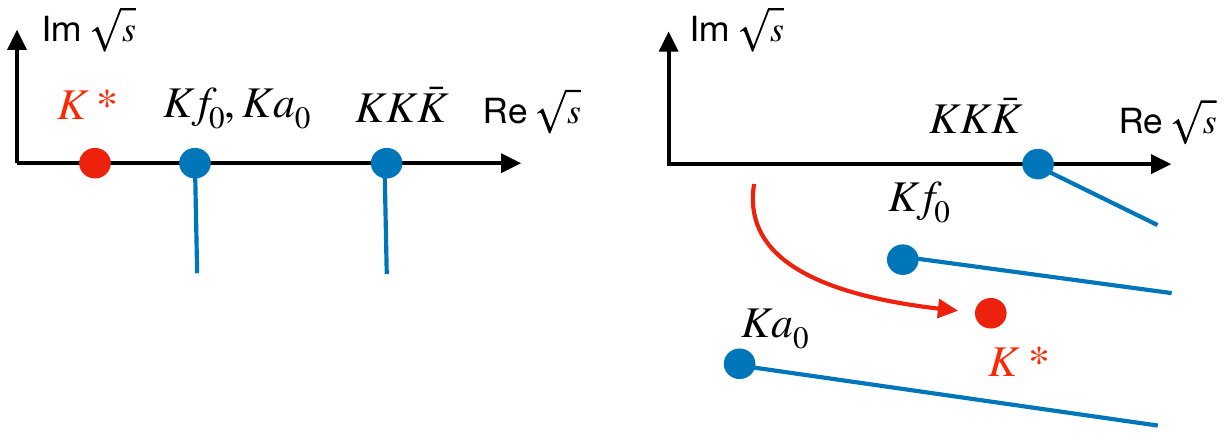}
\end{center}
\caption{Left: Thresholds and their cuts (blue dots and lines) and the $K^*$ bound state (red) for mass-degenerate, stable $f_0$ and $a_0$ isobars, corresponding to the situation quoted in Table~\ref{tab:zerolim}. Right: Same, but for realistic $f_0$, $a_0$ isobars.}
\label{fig:thresholds}
\end{figure}

The analytical continuation of the two-body $t$-matrices obtained by solving Eq.~(\ref{BSon}) to the complex $\sigma$-plane can be done by replacing the $G_i$ of Eq.~(\ref{Gi}) by
\begin{align}
G^C_i=\left\{\begin{array}{l}G_i+\frac{2i\,p_i}{8\pi\sqrt{\sigma}}~\text{for arg}(\sigma-(m_{1i}+m_{2i})^2)>\theta\wedge\text{ Im}\sqrt{\sigma}<0,\\G_i,~\text{otherwise,}\end{array}\right.
\label{eq:rotatecut}
\end{align}
with $\theta\in[-\pi,\pi]$. The standard choice is $\theta=-\pi/2$ for which the cut runs from threshold in the negative Im~$\sigma$ direction. However, as motivated in Sec.~\ref{sec:continuation}, we allow the two-body cuts to rotate to inspect hidden sheets of the three-body amplitude.
Poles related to the scalar resonances $f_0(500)$ [or $\sigma$], $f_0(980)$, $a_0(980)$, and $K^*_0(700)$ (or $\kappa$) are shown in Table~\ref{poles}, indicating the relevance of the meson-meson dynamics to describe the properties of these states. It is important to state here the role of the coupled channel dynamics in the formation of these poles. While $f_0(980)$ can be generated just from $K\bar K$ dynamics as a bound state, the $\pi\pi$ channel is necessary to obtain a finite width. This is in contrast to the case of $a_0(980)$, for which the $\pi\eta$ dynamics is needed to generate it. However, in several approaches, when studying three-body systems involving the $f_0(980)$ and $a_0(980)$ states and investigating the possible generation of resonances, it is common to consider both $f_0(980)$ and $a_0(980)$ as stable $K\bar K$ bound states with the same mass (see Fig.~\ref{fig:thresholds} to the left). 

To be able to take this limit, the inelastic $K\bar K\to \pi\pi,\,\pi\eta$ transitions when solving Eq.~(\ref{BSon}) and their effects on the generation of poles in the three-body system have been studied in this work by replacing $V_{ji}$, with $i\neq j$, by $x V_{ji}$, with $x$ ranging from 0 to 1. The value $x=0$ corresponds to the case in which $f_0(980)$ and $a_0(980)$ are described as $K\bar K$ bound states with the same mass, and $x=1$ represents the full coupled channel calculation, leading to the pole positions shown in Table~\ref{poles}. When varying the parameter $x$, the subtraction constants are also modified accordingly to adjust the real part of the pole positions for $f_0(980)$ and $a_0(980)$ to similar values as those shown in Table~\ref{poles}. The kaon weak decay constant is also slightly modified to adjust the position of the $a_0(980)$ pole to that of the $f_0(980)$ when neglecting the $\pi\eta$ interaction. In this case, the pole position obtained for both $f_0(980)$ and $a_0(980)$ is $\simeq 981.6-i 0$ MeV.  Finally, the $KK$ (strangeness $S=2$) interaction has been described using the lowest-order chiral Lagrangian, finding repulsion and, thus, no pole is generated in this system.
\begin{table}[htb]
\caption{Pole positions (mass $M_R$, width $\Gamma_R$) of the scalar resonances in the isobars.}
\label{poles}
 \begin{tabular}{l|rl}
 \hline\hline
 \text{State}& $(M_R$ & $-i\,\Gamma_R/2)/$~MeV\\
 \hline
  $f_0(500)$& $471.97$&$-i\, 189.1$\\
  $f_0(980)$& $988.08$&$-i\, 17.36$ \\
  $a_0(980)$& $982.32$&$-i\,53.63$ \\
  $K^*_0(700)$& $832.57$&$-i \,230.7$ \\
  \hline\hline
 \end{tabular}   
\end{table}

\begin{figure}[htb]
\begin{center}
\includegraphics[width=0.99\textwidth]{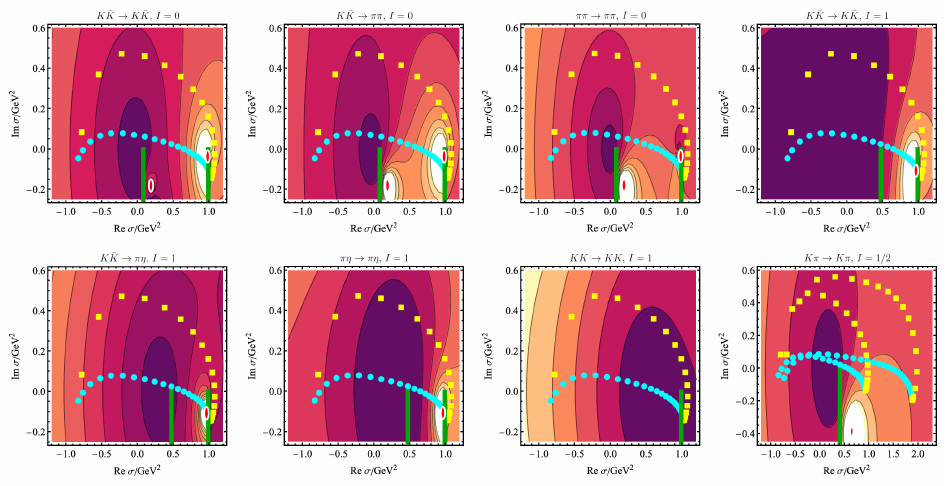}
\end{center}
\caption{Isobar amplitudes $|\tilde{\tau}_{ji}|^2$ from small (dark) to large (white) values for $\sqrt{s}=(1510-70\,i)$~MeV. Poles are indicated with red dots, cuts with green lines, and the SMCs $C_1$ ($C_2$) from Fig.~\ref{fig:demo}, mapped to the $\sigma$ planes using Eq.~\eqref{eq:sigma}, with the yellow squares (turquoise circles). 
For the $K\pi$ isobar there are two different spectator masses ($\pi$ and $\eta$), i.e., two different mappings.
The standard choice for the two-body cuts is shown: they run from their respective threshold in the negative Im~$\sigma$ direction, i.e., at  an angle of $\theta=-\pi/2$ measured from the real axis.}
\label{fig:twosub}
\end{figure}

In Fig.~\ref{fig:twosub} we show the isobar amplitudes (in our case, two-body $t$-matrices) together with typical SMCs mapped to the different $\sigma$ planes using \cref{eq:sigma} at the physical $x=1$. The $f_0(500)/\sigma$, $f(980)$, $a_0(980)$ and $K_0^{*}(700)/\kappa$ are shown with the red dots. It is important to check the entire $\sigma$ plane for problems such as poles on the first Riemann sheet, poles at negative $\sigma$, or non-analyticities from an insufficient parametrization of the two-body amplitudes. Such structures can generate false imaginary parts violating unitarity and/or invalidate the deformation of the SMC into the complex plane for the numerical solution of the scattering equation discussed in the next section. Note that left-hand cuts would require special attention, but the current isobars from the LO chiral interaction do not contain any. 

\subsection{Numerical solution of the scattering problem}
\label{sec:numerics}
Numerical solution techniques for the isobar-spectator $T$-matrix in Eq.~\eqref{eq:TLL} and the production processes $\tilde \Gamma$ and $\breve\Gamma$ were discussed in Refs.~\cite{Feng:2024wyg, Sakthivasan:2024uwd} and reviewed in Ref.~\cite{Doring:2025sgb}. 
The first step consists in discretizing the SMC;
the scattering equation~\eqref{eq:TLL} becomes a matrix equation that reads 
\begin{align}
    T=\tilde B+\tilde B \tilde\tau XT,\quad X=\text{diag}\left(\frac{\diff p_l\,p_l^2}{(2\pi)^3\,2E^{(1)}_{p_l}},\dots,
    \frac{\diff p_l\,p_l^2}{(2\pi)^3\,2E^{(N)}_{p_l}}\right) \ ,
\end{align}
where the total energy argument was omitted. The absence of  indices implies  vectors ($\tilde\Gamma$, $\breve\Gamma$, $D,\,\breve v$) and matrices ($T$, $\tilde B$, $\tilde\tau$, $X$) in the discussed space of channels, isobar sub-channels, and discretized spectator momenta $p_l$. The $\diff p_l$ are the integration weights. The energies in the kinematic factor $X$  depend on the spectator masses in the $N=7$ channels.
Similarly,
\begin{align}
   \tilde \Gamma= T\tilde\tau XD,\quad 
   \breve\Gamma=\breve v\tilde \tau\left(\tilde\Gamma+D\right) \ .
\end{align}
The latter quantity can be written as
\begin{align}
\breve\Gamma=\breve v\left(\tilde\tau^{-1}-\tilde BX\right)^{-1}D \ .
\label{eq:gammabrevnum}
\end{align}

The SMC can be deformed from the real axis into the lower momentum half-plane which respects the $+i\epsilon$ prescription in $\tilde B$ and avoids the three-body cuts in $\tilde B$. In addition, the SMC mapped to the isobar sub-energy plane $\sigma$ using Eq.~\eqref{eq:sigma} is situated in the upper $\sigma$ half plane, i.e., it lies entirely on the first Riemann sheet avoiding branch cuts from isobar coupled channels as \cref{fig:twosub} shows (adapted to the case of real $\sqrt{s}$ which would shift the shown SMCs upwards). See Ref.~\cite{Sadasivan:2021emk} for a detailed and illustrated discussion. 

An additional challenge concerns the calculation of the amplitude at real-valued spectator momenta. 
For this, in Ref.~\cite{Feng:2024wyg} the contour deformation method of Cahill and Sloane (CS)~\cite{Cahill:1971ddy} was tested and found to lead to identical results as the method by Schmidt and Ziegelmann~\cite{schmid1974quantum} referred to as ``direct inversion''. In contrast to other methods, the CS method is inherently numerically stable,\ but complicated to implement in the presence of multiple channels with different masses. The direct-inversion method does not require any contour deformation and the problems with singular integrands (of logarithmic and/or simple-pole type) are moved from the level of the integral equation to the level of integrals, where they are much easier to deal with. 

A third method was proposed in Ref.~\cite{Sakthivasan:2024uwd} using continued fractions to extrapolate from the complex SMC, at which the solution is known, to real, outgoing spectator momenta. The idea is the same as in Ref.~\cite{Sadasivan:2020syi}, but continued fractions are more stable and capable of even extrapolating to sharp threshold cusps. This makes it the method of choice for the present work, due to its very simple implementation and independence of channel masses. 
\begin{figure}[htb]
\begin{center}
\includegraphics[width=0.99\textwidth]{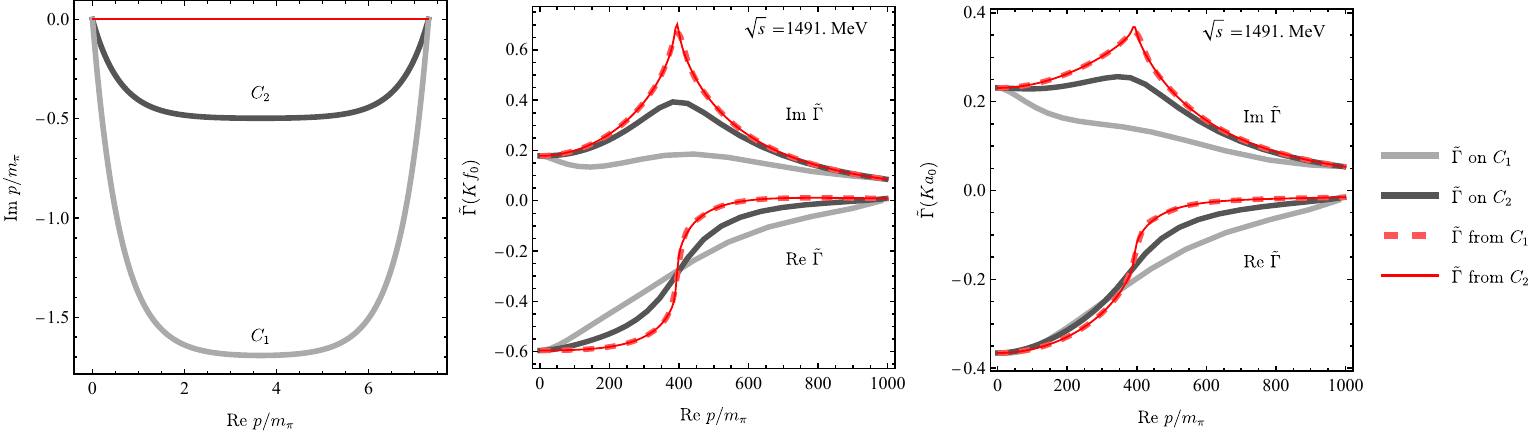}
\end{center}
\caption{Analytic continuation of the production amplitude $\tilde\Gamma$ to real momenta. Left: Two different SMCs $C_1$ and $C_2$. Center and right: $\tilde\Gamma$ for the two channels of a coupled $Kf_0,\,Ka_0$ channel model. The figures contain both real (upper three curves in each plot) and imaginary parts (lower three curves in each plot). The extrapolations from $C_1$ and $C_2$ to real spectator momenta using continued fractions is shown with the red dashed and the red solid lines, respectively. }
\label{fig:demo}
\end{figure}
The method is demonstrated in Fig.~\ref{fig:demo}. The production amplitude $\tilde\Gamma$ from Eq.~\eqref{eq:normalgamma} in a reduced, two-channel $Kf_0,\,Ka_0$ model is evaluated with two different SMCs $C_1$ and $C_2$ parametrized as~\cite{Sadasivan:2021emk} 
\begin{align}
p=t+iV_0\left(1-e^{-t/w_a}\right)
\left(1-e^{-(t-\Lambda)/w_a}\right),\quad t\in[0,\Lambda] \ .
\label{eq:smc}
\end{align}
The extrapolations from these SMCs, shown with the red lines, agree well for a typical choice of 36 spectator-momentum Gauss points, and both reproduce the sharp $K\bar K$ cusp from the $f_0,\,a_0$ isobars, even the extrapolation from the more distant SMC $C_1$.  

\subsection{Analytic continuation}
\label{sec:continuation}
For the search of resonance poles in the lower  $\sqrt{s}$ half plane one can continue working with a complex SMC ensuring that the contour never crosses any of the cuts in the three- and two-body amplitudes. This requires additional consideration as the position of these cuts depends on $\sqrt{s}$ and its imaginary part, as well as the SMC mapped to the sub-energy $\sigma$ plane. This is discussed in detail in Ref.~\cite{Sadasivan:2021emk}, building on previous work along the same ideas~\cite{Doring:2009yv}. The main aspects of analytic continuation for the present framework (and many related frameworks) have been reviewed in Ref.~\cite{Doring:2025sgb}. Here we discuss the present kinematic situation and add an improved discussion of allowed and forbidden regions of the $\sqrt{s}$-plane for the search of resonance poles.

\begin{figure}[htb]
\begin{center}
\includegraphics[width=0.49\textwidth]{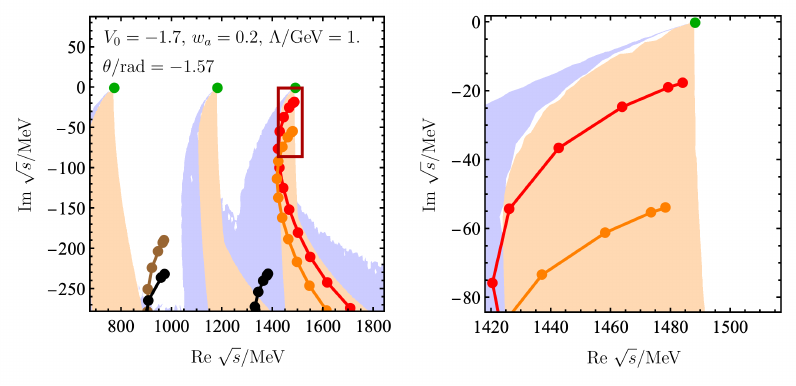}
\includegraphics[width=0.49\textwidth]{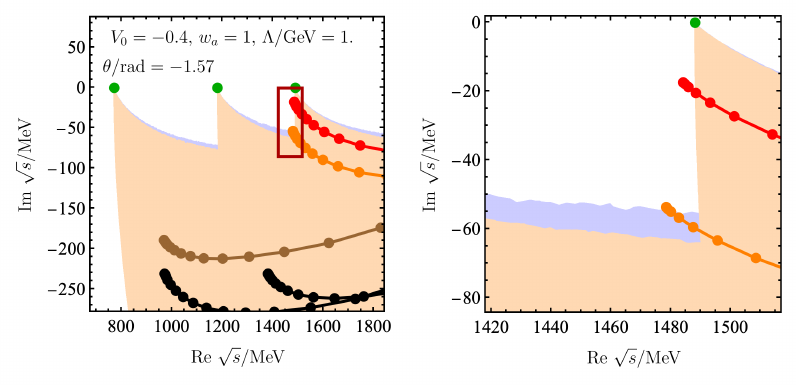}
\end{center}
\caption{Three-body cuts and isobar-resonance cuts in the complex plane for the SMC reaching far into the complex plane (left two pictures) or not so far into the complex plane (right two pictures). The second and fourth pictures zoom into the rectangle shown in the first and third picture, respectively. The lines with dots show the $f_0(980)$ cut (red), $a_0(980)$ cut (orange), $\sigma$ cut (brown), and $\kappa$ cuts (black). There are two cuts from the $\kappa$ due to the different spectator masses according to $\pi\kappa$ and $\eta \kappa$. The blue shaded area shows the regions of three-body singularities from the forward-going particle-exchange process corresponding to the different thresholds $\pi\pi K$, $\pi\eta K$, and $KK\bar K$, indicated with the green dots. The orange shaded area is an excluded zone when the SMC hits two-body cuts in the isobar subsystems, that run from the two-body thresholds into the negative Im $\sigma$ direction (see Fig.~\ref{fig:twosub}). The corresponding rotation angle of $\theta=-\pi/2\approx -1.57$ according to \cref{eq:rotatecut} is indicated, together with chosen parameters for the SMC according to Eq.~\eqref{eq:smc}.}
\label{fig:forbidden}
\end{figure}

Depending on the SMC and choice of the cuts in the isobar amplitudes, some regions in the complex $\sqrt{s}$-plane are excluded from pole searches as indicated in Fig.~\ref{fig:forbidden}.
The parameters quoted in the picture correspond to Eq.~\eqref{eq:smc}.
Forbidden regions arise, on one hand, from three-body cuts in the forward-going interaction terms of Eq.~\eqref{btilde}, shown as the blue regions (we omit an analogous discussion for the backward going parts of $\tilde B$). They touch the physical $\sqrt{s}$ axis at the respective three-body thresholds $\pi\pi K$, $\pi\eta K$, and $KK\bar K$. There is the three-body cut running inside these forbidden regions from these thresholds in the $-\text{Im}\sqrt{s}$ direction~\cite{Doring:2009yv}, shown symbolically in \cref{fig:thresholds}.

Other cuts arise from the two-body input $\tilde\tau$~\cite{Doring:2009yv}; if an isobar contains a resonance, there are complex thresholds at $\sqrt{s}=m_i+R_i$, where $m_i$ is the spectator mass and $R_i$ is the complex pole position in the variable $\sqrt{\sigma}$. Properties of the complex branch points were determined in Ref.~\cite{Ceci:2011ae}. The shape of the cuts associated with these branch points depends on the chosen SMC. 
They are indicated in the figure with different colors (see the caption). The dots indicate the position of specific Gau{\ss} points of the SMC. Due to the numerical discretization, these cuts appear as a series of poles rather than a discontinuity,  see also Fig.~9 in Ref.~\cite{Sadasivan:2021emk} and the center panel of \cref{fig:trajectories} below. 

\begin{figure}[htb]
\begin{center}
\includegraphics[width=0.49\textwidth]{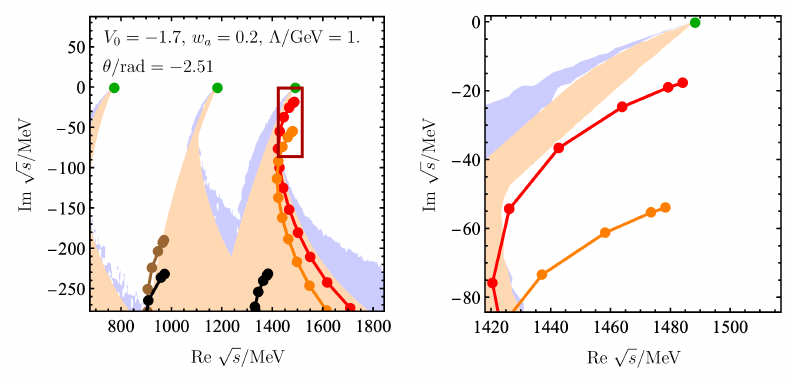}
\includegraphics[width=0.49\textwidth]{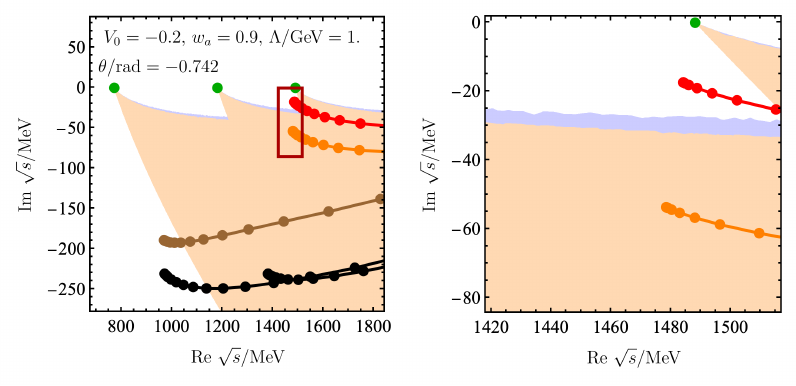}
\end{center}
\caption{Three-body cuts and isobar-resonance cuts in the complex plane similar to Fig.~\ref{fig:forbidden}. The only difference lies in the two-body cuts (thick green lines in Fig.~\ref{fig:twosub}) that are rotated further to the left (left panel) or to the right (right panel) according to the quoted rotation angle $\theta$ that is measured from the Re~$\sigma$ axis.}
\label{fig:forbidden2}
\end{figure}

A second excluded zone in the complex $\sqrt{s}$-plane arises from a geometric restriction illustrated in Fig.~\ref{fig:twosub}. The figure shows the shallow (turquoise) and  deep (yellow) SMCs $C_2$ and $C_1$ mapped from Fig.~\ref{fig:demo} to the sub-energy $\sigma$ plane.
The shallow path crosses the $K\bar K$ two-body cut (thick green line) and, therefore, cannot be used at the chosen complex $\sqrt{s}$. This and similar path crossings lead to excluded areas shown in orange in Fig.~\ref{fig:forbidden}. 
Besides contour deformations to access forbidden areas one can also rotate the two-body cuts in $\tilde\tau$ (the green bars in Fig.~\ref{fig:twosub}) to the left (right) according to \cref{eq:rotatecut}, which uncovers parts of the $\sqrt{s}$-plane shown in Fig.~\ref{fig:forbidden2} to the left (right).
This rotation can be used to inspect parts of the complex $\sqrt{s}$ plane that are hidden behind the three-particle thresholds. This is particularly useful to search for virtual states or ``shadow poles''.

\section{Results}
\subsection{Stable, mass-degenerate isobars}
\label{sec:boundstate}
First, we study the limit of stable, mass-degenerate isobars for $f_0$ and $a_0$. For this, the light channels $\pi\pi$ and $\pi\eta$ are decoupled and the $K\bar K$ interaction  strength is modified such that both resonances appear as bound states at $\sqrt{\sigma}\approx 981.6$~MeV, i.e., around 10 MeV below the $K\bar K$ threshold, as described in ~\cref{sec:isobars}. This corresponds to $x=0$ for the tuning parameter. 

In this limit, indeed, a bound state is obtained at the binding energies shown in Table~\ref{tab:zerolim}. In this stable-isobar limit, the only channels are $Kf_0$ and $Ka_0$ (first two columns quoting binding energies), and possibly also the $KK_2$ channel (last column).
The kinematic situation is illustrated in Fig.~\ref{fig:thresholds} to the left. 
The binding energy is measured with respect to the $Kf_0$ (or $Ka_0$) threshold.
\begin{table}[htb]
\center
    \begin{tabular}{c|ccc}
    \hline\hline
    $\Lambda$/GeV
   & $W_b$/MeV (w/o bw) & 
    $W_b$/MeV (w b)& 
    $W_b$/MeV (w $K_2$)   \TT\BBB \\
    \hline
0.8 &1.13&1.42& 0.60     \TT \\
1.0 &1.18&1.44& 0.63\\
1.2 &1.22&1.45& 0.63
\BBB\\
\hline
\hline
    \end{tabular}
        \caption{Binding energy $W_b$ of the $Kf_0/Ka_0$ bound state for different cutoffs $\Lambda$. The columns indicate the case without (w/o~b) and with (w~b) backward going exchanged particle contribution, as well as the case when including the strangeness$-$2 $KK-$isobar (w~$K_2$) .}
    \label{tab:zerolim}
\end{table}
The results are in line with those determined in the study of Ref.~\cite{Zhang:2021hcl}, which considers not only backward-going exchange kaons in their TOPT formalism, but also backward propagating kaons in the $s$-channel, and kaon renormalization. Here, we only consider a subset of these possibilities. Another difference lies in the representation of the amplitude. In Ref.~\cite{Zhang:2021hcl}, isobar fields are used while here we formulate the interaction through contact terms; bound states/resonance emerge in the unitarization. In spite of these differences, the very small binding energy of around 1 MeV is remarkably close to the one of Ref.~\cite{Zhang:2021hcl} although, in detail, results are difficult to compare. In both calculations, the cutoff dependence of the result is small. Notably, no bound state is obtained in either channel for the case in which both channels are decoupled, making this a truly coupled-channel bound state. 

\subsection{Results for the 7-channel model}

\begin{figure}[htb]
\begin{center}
\includegraphics[width=1.\textwidth]{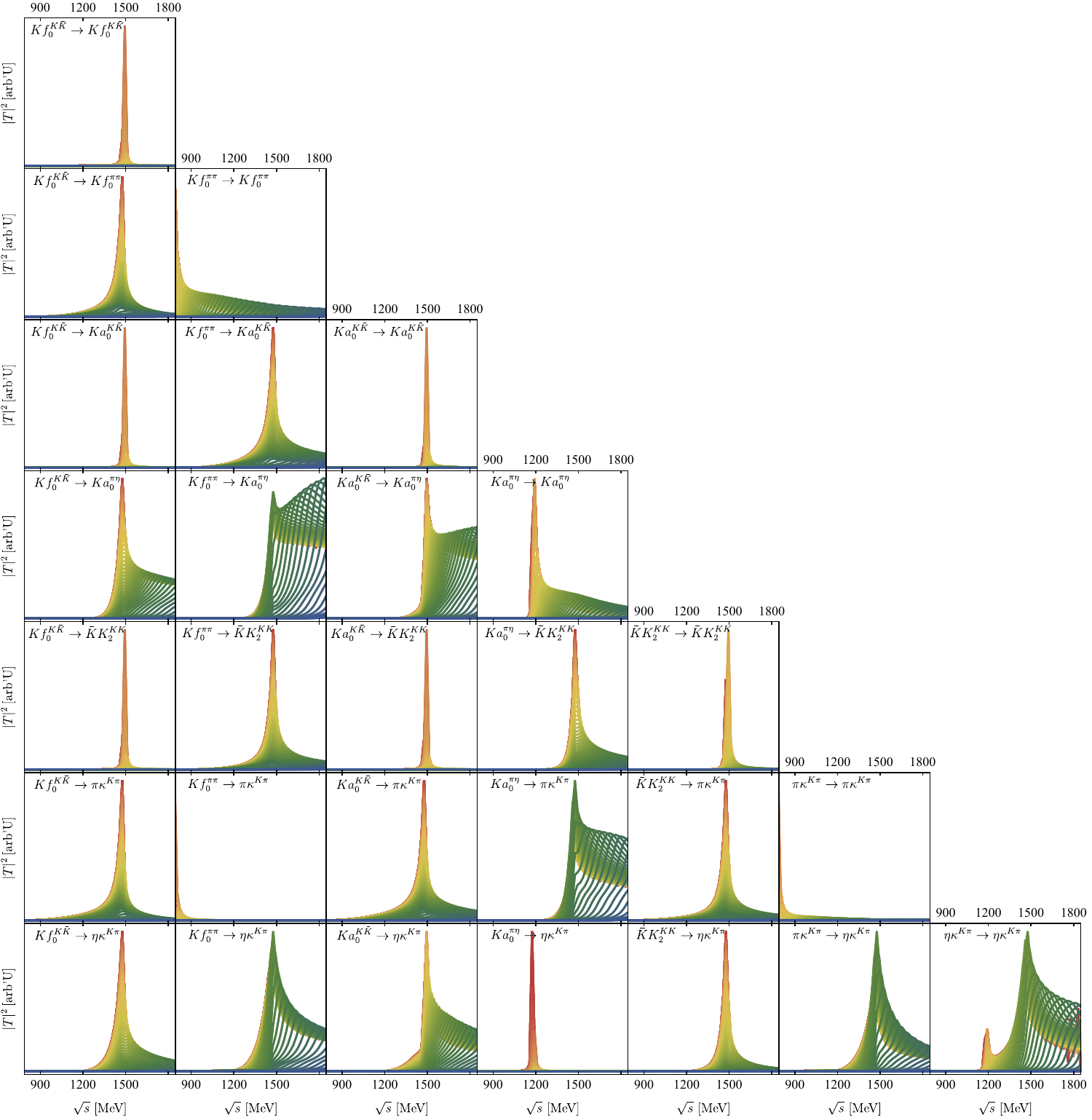}
\end{center}
\caption{
The $T$-matrix $|T_{ji}(\sqrt{s},p,p)|^2$ for channel transitions $j\leftarrow i$ and pseudo-real spectator momenta $p-i\Delta$ from small (red) to large (blue) values for $p$, for $\Delta\approx 30$~MeV.}
\label{fig:tmat}
\end{figure}
The $T$-matrix for the full coupled-channel system is shown in Fig.~\ref{fig:tmat} for different spectator momenta as described in the caption. We observe sharp structures close to the $KK\bar K$ threshold at $\sqrt{s}\approx 1.49$~GeV due to on-shell kaon exchange in processes like $Kf_0^{K\bar K}\to Kf_0^{K\bar K}$.
The enhancement comes from the partial-wave (PW) projection of the forward-going part of $\tilde B$ from \cref{btilde}. The physical region (above threshold) for $p=p'$ is bound through $x=\cos\theta\in[-1,1]$. At both $x=\pm 1$ one has logarithmic singularities in the PW projected $\tilde B$ in the variable $\sqrt{s}$ for fixed $p=p'$, say $\sqrt{s_+}$ and $\sqrt{s_-}$. As $\sqrt{s}$ approaches the threshold $\sqrt{s_0}$ from above, $\sqrt{s_+}\to \sqrt{s_-}\to \sqrt{s_0}$, the logarithmic singularities fall together at threshold, enhancing each other while  Im~$\tilde B(s,p',p)$ diverges. This is a well-known kinematic effect and not directly related to triangle singularities.
Through coupled-channel effects this enhancement is also visible in other transitions, like $Kf_0^{\pi\pi}\to Ka_0^{\pi\eta}$ for which neither in the first or the last transition a kaon is exchanged. For $Ka_0^{\pi\eta}\to Ka_0^{\pi\eta}$, $Ka_0^{\pi\eta}\to \eta \kappa^{K\pi}$, and $\eta \kappa^{K\pi}\to \eta \kappa^{K\pi}$ one observes an enhancement at $\sqrt{s}\approx 1.18$~GeV, due to a pion exchange that can go on-shell in the $Ka_0^{\pi\eta}\to \eta \kappa^{K\pi}$ transition right at the $K\pi\eta$ threshold. 
In other words, the $\tilde B$-exchange process exhibits a logarithmic singularity when the $KK\bar K$ state in the exchange $\tilde B$ goes on-shell at the $KK\bar K$ threshold, or the $K\pi\eta$ state in $\tilde B$ can go on-shell  at the $K\pi\eta$ threshold. In both cases $\tilde B\approx  T$ for the corresponding channel transition.  

\begin{figure}[htb]
\begin{center}
\includegraphics[width=0.95\textwidth]{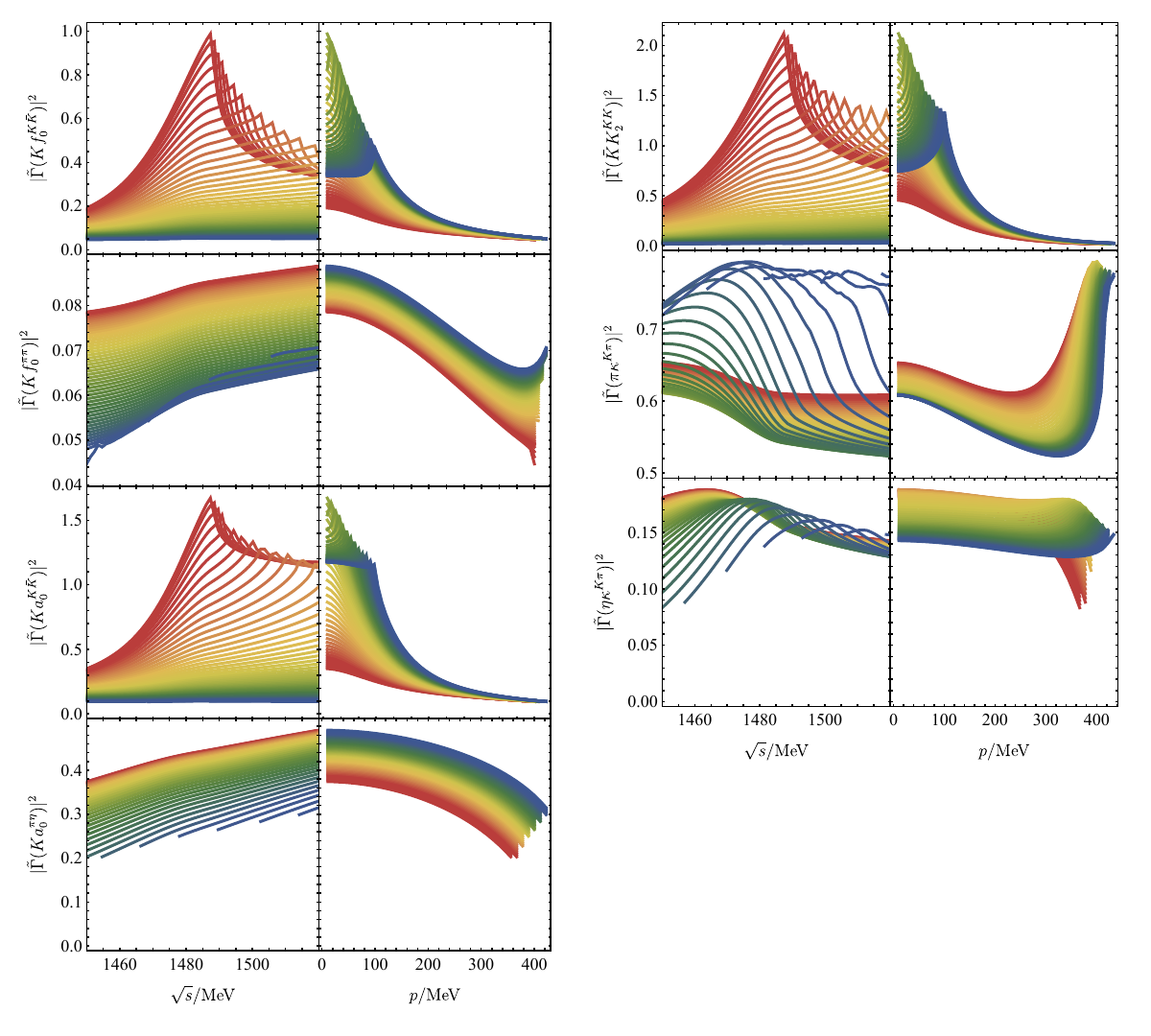}
\end{center}
\caption{
Connected production amplitude, $|\tilde\Gamma|^2$, for the 7-channel model, as a function of three-body energy (left row) and spectator momentum (right row). The color coding for the different lines is from smaller to larger values for spectator momenta (left row) and energy (right row). For the channels 2, 4, 6, and 7, we only show the amplitude in the physical region, while for channels 1, 3, and 5 the amplitude is shown also in the unphysical region below the $KK\bar K$ threshold to demonstrate the moving cusp, further illustrated in Fig.~\ref{fig:mov_cusp}.}
\label{fig:gammatilde}
\end{figure}
In Fig.~\ref{fig:gammatilde} the connected production amplitude $\tilde\Gamma$ is shown for the seven channels and real spectator momenta using the continued fraction method from \cref{sec:numerics}. For some of the channels (2,4,6,7) only amplitudes in the physical region are displayed. As that region depends on the spectator momenta, some of the curves start/end inside the shown energy window. For channels 1, 3, and 5, we also show the amplitude below the physical window because these outgoing states show a moving cusp at the $KK\bar K$ threshold further discussed below. This cusp is prominently enhanced for small spectator momenta. We attribute this to a triangle singularity, discussed below. 

Part of the enhancement is also attributed to the presence of the complex $Kf_0$ threshold right at $\sqrt{s}\approx 1.48$~GeV, as shown in Fig.~\ref{fig:forbidden} with the red symbols. Complex threshold openings can in some cases indeed lead to misidentification as resonances~\cite{Ceci:2011ae}.

One could wonder why there is also a strongly enhanced cusp for channel 5, $\bar K K_2$. After all, the repulsive $K_2$ subchannel does not contain a narrow resonance (ruling out triangle singularities) or complex threshold opening. We attribute the appearance of the enhanced cusp to its presence in other channels (1 and 3), in combination with coupled-channel effects and the fact that the $K_2$ channel does have a $KK$ cusp just like the $f_0$ and $a_0$ subchannels.

If one plots the last channel $\tilde \Gamma (\eta \kappa^{K\pi})$ for  $\sqrt{s}\approx 1.18$~GeV (not shown), one does \emph{not} observe any enhanced cusp structure at the $K\pi\eta$ threshold, even though there is one visible in the $T$-matrix, as discussed before (see Fig.~\ref{fig:tmat}). While there is no triangle singularity for this kinematics, one can still wonder what happened to the enhancement  observed for $T$. In this case, it disappears because $\tilde\Gamma$ contains $T$ in half-off-shell kinematics, i.e., there is an integration over the incoming momentum for $T$  according to \cref{eq:normalgamma}.

\begin{figure}[htb]
\begin{center}
\includegraphics[width=0.4\textwidth]{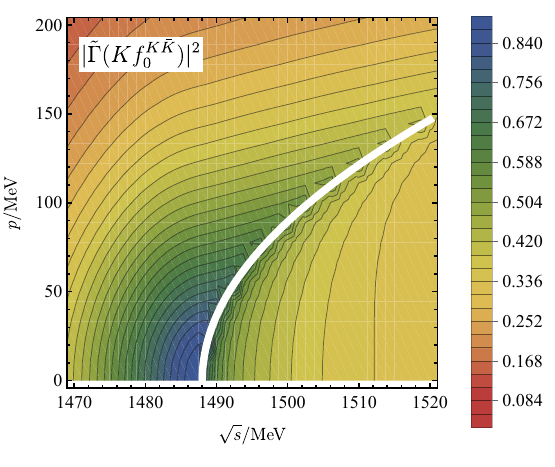}
\end{center}
\caption{Production amplitude $|\tilde\Gamma|$ for the $Kf_0^{K\bar K}$ channel as a function of three-body energy $\sqrt{s}$ and spectator momentum $p$. There is a cusp that follows the white line given by Eq.~\eqref{eq:plim} that marks the boundary between the physical region (below the line) and the unphysical region (above the line), separated by the moving cusp.}
\label{fig:mov_cusp}
\end{figure}
Returning to the discussion of the moving cusp in $\tilde\Gamma$ for channels 1, 3, and 5, \cref{fig:mov_cusp} shows the size of $\tilde\Gamma$ for channel 1. Indeed, the position of the cusp (white line) depends on the spectator momentum. In fact, the cusp simply indicates the limits of the physical region for the $KK\bar K$ states given by~\cite{Sadasivan:2021emk}
\begin{align}
    p=\frac{\sqrt{9m_K^4-10m_K^2s+s^2}}{2\sqrt{s}} \ .
    \label{eq:plim}
\end{align}

\begin{figure}[htb]
\begin{center}
\includegraphics[width=1.\textwidth]{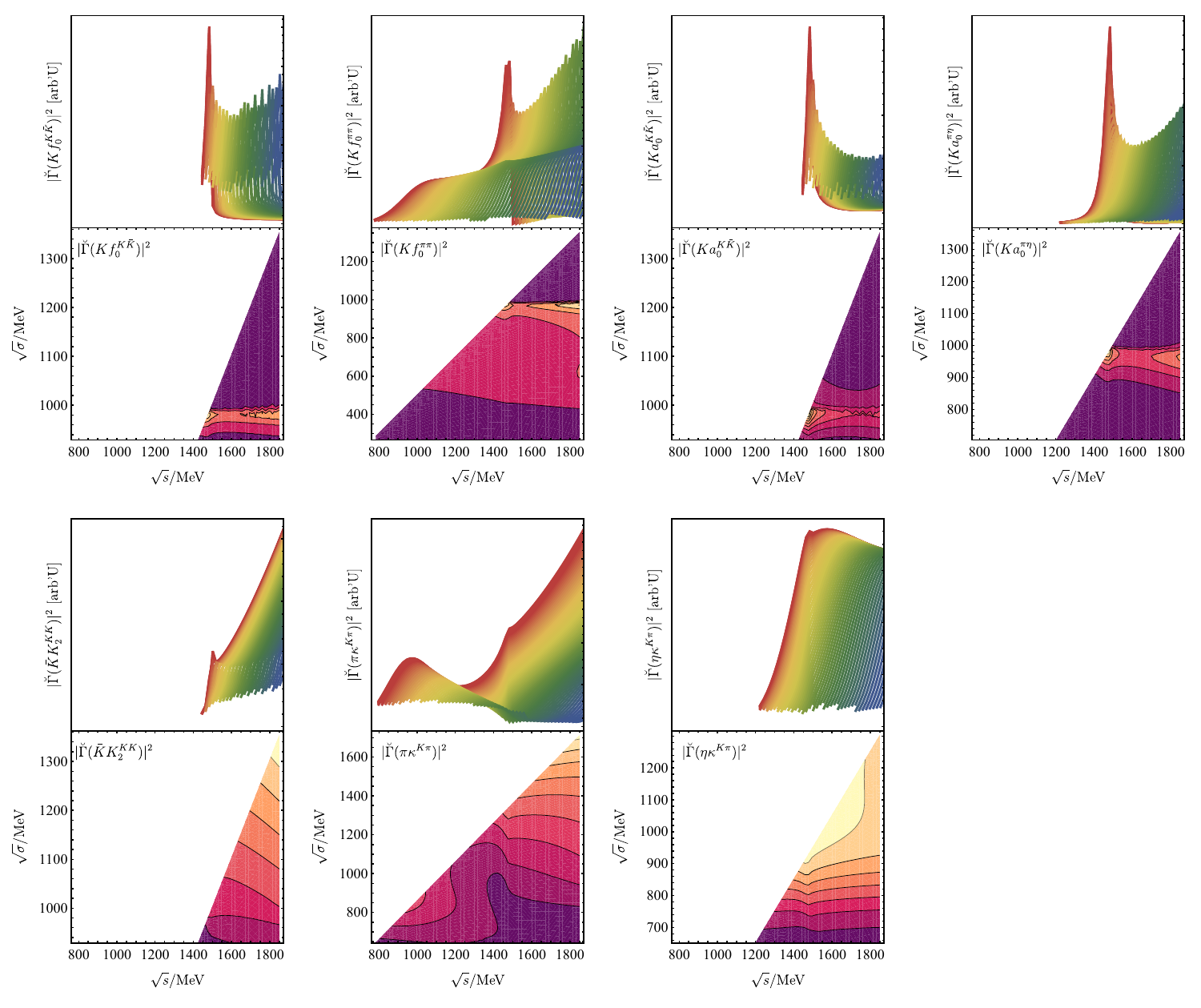}
\end{center}
\caption{
Full production amplitude, $|\breve\Gamma|^2$, for the 7-channel model. The upper plots show the amplitudes-squared for small (red) to large (blue) spectator momenta. Only physical spectator momenta are shown, which leads to a slightly choppy appearance due to finite energy resolution of the plots. }
\label{fig:gammabreve}
\end{figure}
In Fig.~\ref{fig:gammabreve}, the full production amplitude $\breve\Gamma$ from \cref{eq:gammabrev} (or \cref{eq:gammabrevnum}) is shown. The $\pi\pi K$ and $\pi\eta K$ decay channels 2 and 4 prominently show the enhanced $KK\bar K$ cusp at around $\sqrt{\sigma}\approx 1480$~MeV, in contrast to the structureless channels 2 and 4 of $|\tilde\Gamma|^2$ shown in Fig.~\ref{fig:gammatilde}. This originates from the underlying enhancement at this energy in the $Kf_0^{K\bar K}$ channel, followed by a sub-channel $K\bar K\to\pi\pi$ transition, leading to a physically detectable signal in channel 2. 

Similarly, the strong signal in channel 4 at $\sqrt{s}\approx 1.48$~GeV arises from the structure in channel 3, followed by a $K\bar K\to\pi\eta$ transition. As a result, the ``resonance signal'' should be detectable for the $K\pi\eta$ final state. Finally, we observe a very broad bump in channel 6, at the lower end of the physical region but well within it for small spectator momenta, which should make it detectable in experiment. We attribute that to the complex threshold/branch point of the $\pi\kappa$ amplitude (see the black dots in \cref{fig:forbidden} at $\sqrt{s}\approx 1$~GeV). We do not find any dynamically generated pole close to that complex $\pi\kappa$ threshold, but have not made any attempts to isolate a possible pole by taking a zero-width limit for the $\kappa$ resonance the way it is done for the $Kf_0/Ka_0$ case discussed before.

\begin{figure}[htb]
\begin{center}
\includegraphics[width=1.\textwidth]{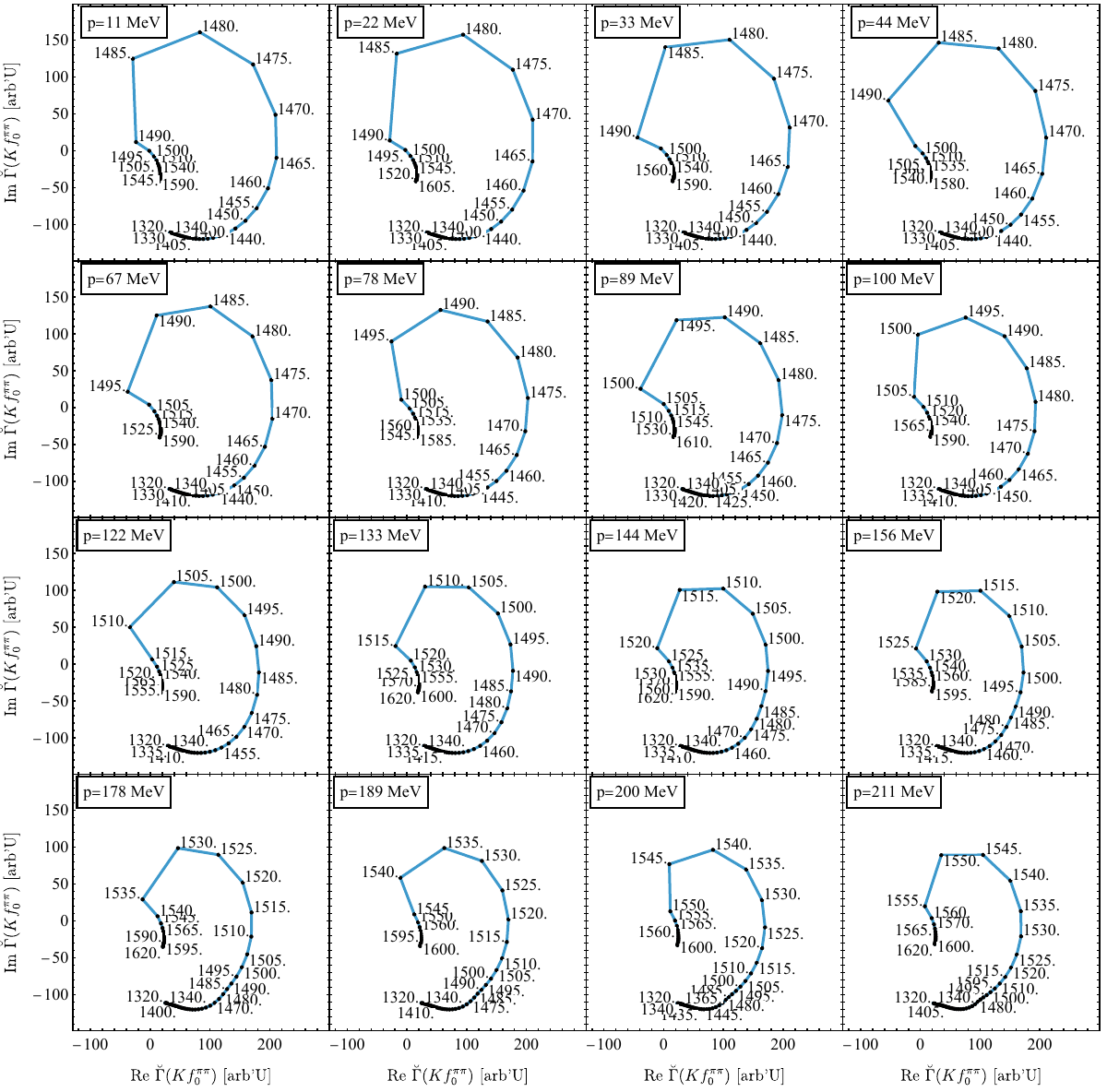}
\end{center}
\caption{
``Argand plot'' of the production amplitude $\breve\Gamma$ for channel 2 with $K\pi\pi$ final state, around the peak position of the bump at $\sqrt{s}\approx 1.48$~GeV shown in \cref{fig:gammabreve}. The fixed spectator momenta are shown in the upper left corner of each plot. The labels indicate the three-body energy $\sqrt{s}$ in MeV. Note that this quantity cannot be directly compared to the Argand plots in Ref.~\cite{LHCb:2017swu} as explained in the text.}
\label{fig:argand}
\end{figure}

While none of the discussed bumps in $\breve\Gamma$ are directly related to resonances, they could easily be mistaken for one.
In this context, one can study the phase motion of the structure in channel 2 for different spectator momenta. This is shown in \cref{fig:argand}.
While not a perfect circle, there is undeniably a strong phase motion that could be erroneously taken as a resonance signal. Although it is tempting to compare the predicted Argand plots to the experimental one from Ref.~\cite{LHCb:2017swu}, it is unclear if such a comparison makes sense. Experimental isobar models usually rely on a factorization of the three-body part (the complex-valued amplitude shown in their Argand plots), multiplied with a final-state isobar. In the present formulation, we also have a factorized, final isobar in $\breve\Gamma$, but the three-body part $\tilde\Gamma$ does not only depend on the three-body energy but also on the sub-energies in a non-factorizable way as required by unitarity.

\subsection{Pole trajectory}
\label{sec:poletrajectory}
\begin{figure}[htb]
\begin{center}
\includegraphics[width=0.2499\textwidth]{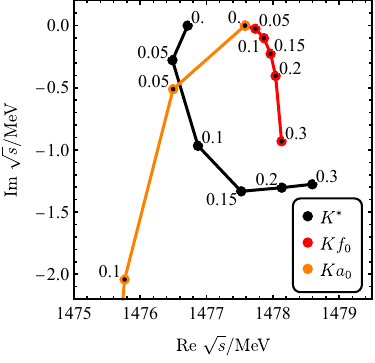}
\includegraphics[width=0.2499\textwidth]{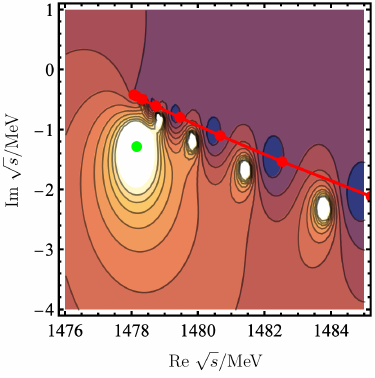}
\includegraphics[width=0.34\textwidth]{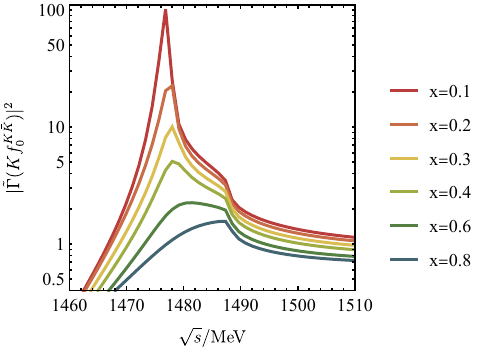}
\end{center}
\caption{
{\bf Left}: Trajectories of the dynamically generated $K^*$ pole (black), the $Kf_0(980)$ complex threshold (red), and the $Ka_0(980)$ threshold (orange) as a function of a parameter $x\in [0,0.3]$ with limits $x=0$ (stable, equal-mass $f_0$, $a_0$) and $x=1$ (physical $f_0,\,a_0$). For $x>0.3$, the $K^*$ pole disappears behind the $Kf_0(980)$ threshold and is no longer traced. {\bf Center}: The $|T(Kf_0^{K\bar K}\to Kf_0^{K\bar K})|^2$ transition with smaller (larger) magnitude in darker (lighter) colors, for $x=0.2$. The $K^*$ pole (green dot) is clearly distinct from the set of poles associated with the $K f_0$ branch cut (small white dots), which is rotated to the right corresponding to the right two panels in Fig.~\ref{fig:forbidden2}.
The figure illustrates the numerical difficulty to separate the pole from the nearby cut. {\bf Right}: Production amplitude for channel 1 at a small spectator momentum of $p=11$~MeV for different $x$.
}
\label{fig:trajectories}
\end{figure}

After discussing the formation of a bound state in the limit of stable, mass degenerate $f_0$ and $a_0$ isobars in Sec.~\ref{sec:boundstate}, the question remains where this state moves once all seven channels are switched on. No direct evidence of it is visible in \cref{fig:gammatilde}. To be able to trace the pole, the light channels in the  $f_0$ and $a_0$ isobars, i.e., $\pi\pi$ and $\pi\eta$, are slowly switched on by tuning a parameter $x$ from 0 to 1 that multiplies the corresponding offdiagonal transition amplitude $V_{ij}$ when calculating the $t$-matrix via Eq.~(\ref{BSon}). The situation $x=0$ ($x=1$) corresponds to the mass-degenerate, stable isobar case (full chiral unitary model with realistic isobar widths) as described in \cref{sec:isobars}. For simplicity, we restrict the model to the three channels $Kf_0$, $Ka_0$, and $\bar KK_2$. The inclusion of the other channels has almost no effect on the pole position. The resonance trajectory is shown qualitatively to the right in \cref{fig:thresholds} in red and quantitatively to the left in \cref{fig:trajectories} in black. 
There, the trajectories of the $Kf_0$ and $Ka_0$ complex thresholds are also displayed. Clearly, the pole is attracted by the $Kf_0$ threshold (red). 

The center panel shows how close the pole (green dot) comes to the mentioned threshold. It also shows the position of the cut that is numerically realized through a series of poles due to the discretization of the spectator momentum. To be able to localize the pole behind the cut one has to rotate the latter as much to the right as possible, which is achieved through a suitable SMC in combination with a rotation of the two-body cuts in the isobars as shown in the right two panels of \cref{fig:forbidden2}.

As the left panel of \cref{fig:trajectories} shows, the pole is effectively shielded from the real axis by the cut (i.e., it is on a hidden Riemann sheet). This effect is expected to continue for larger values of $x>0.3$ for which we lost track of the pole as it got hidden even further behind the complex $Kf_0$ threshold.
The shielding of the pole is further illustrated in the right panel. It shows the production amplitude $\tilde\Gamma$ for channel 1 at the smallest spectator momentum, i.e., when the cusp effect in \cref{fig:mov_cusp} is largest. For the change from $x=0.1$ to $x=0.2$ the amplitude decreases from about 100 to 23, but the imaginary part of the pole only increases from 1~MeV to 1.3~MeV. At $x=0.3$ the amplitude is 10\% of that at $x=0.1$ although the width of the resonance has only increased by 20\%. While for values of $x$ larger than that we lose track of the pole, it does not move substantially further into the complex plane (otherwise we would have found it). Yet, the amplitude reduces to around 2\% of the $x=0.1$ case, leaving only an enhanced $KK\bar K$ cusp and clearly demonstrating the complete shielding of the dynamically generated \ks by the $Kf_0$ cut.

\subsection{A triangle singularity at threshold}
\label{sec:triangle}
Triangle singularities (TSs)~\cite{ Guo:2019twa} correspond to enhancements in the amplitude when several particles go on-shell simultaneously in certain transitions. For example, the $a_1(1460)$ resonance~\cite{COMPASS:2015kdx} has been explained as a triangle singularity~\cite{Mikhasenko:2015oxp, Aceti:2016yeb, COMPASS:2020yhb}: one has a very narrow $K^*(892)$ and a spectator kaon, e.g., $K^*\bar K$, with the $K^*$ emitting an outgoing spectator pion and an interchanged kaon that fuses with the incoming spectator kaon to form a two-body system with invariant mass $m_4$ (e.g., $f_0$). See Refs.~\cite{Guo:2019twa, Liu:2015taa, Debastiani:2018xoi,  Isken:2023xfo, Dai:2018rra, Liang:2019jtr, Jing:2019cbw, Guo:2019qcn, Molina:2020kyu} for more examples and explanations for TSs.

In Ref.~\cite{Bayar:2016ftu}, the kinematic condition of having a triangle singularity was determined, as well. We have checked that their momentum $q_{a-}$ equals the ``triangle momentum'' in Ref.~\cite{Feng:2024wyg}, $P = p_-(m_1,m_3,m_u,-1)$ with the expression for $p_-$ from Eq.~(2.28) of Ref.~\cite{Feng:2024wyg}, $m_1(m_3)$ the incoming (outgoing) spectator and $m_u$ the exchanged mass. In addition, the scattering angle is $\cos\theta=-1$: The exchanged kaon flies in the opposite direction to its mother resonance, the $K^*$, to align with the incoming spectator kaon. The triangle condition is then
\begin{align}
    P=q_{a-}=q_\text{c.m.}(\sqrt{s},m_1,m_2) \ ,
\end{align}
with $q_\text{c.m.}$ being the three-momentum in the three-body center-of-mass-frame of the particle with mass $m_1=m_K$ and the pair with invariant mass $m_2=\sqrt{\sigma_{K\bar K}}$ at total energy $\sqrt{s}$. 

In the current situation one finds this equation to be exactly fulfilled for the transition $K(K\bar K)_{I=0,1}\to K(K\bar K)_{I=0,1}$ at $\sqrt{s}=3m_K$ for $m_2=m_4=2m_K$ and $m_1=m_3=m_u=m_K$. Considering the proximity of the $f_0$ and $a_0$ masses to the $K\bar K$ threshold ($\sqrt{\sigma_{K\bar K}}\approx 992$ MeV) and their finite widths, a triangle diagram involving $K f_0 \bar K$/ $K a_0\bar K$ intermediate states to produce $Kf_0/Ka_0$ final states should give rise to a triangle singularity at $\sqrt{s}\approx 3m_K$. Using the terminology of Ref.~\cite{Bayar:2016ftu}, when $m_2=\sqrt{\sigma_{K\bar K}}=2m_K$, $q_{a+}\to q_{a-}$ and the condition $q_{\text{c.m.}}\to q_{a-}$ produces a triangle singularity at the threshold~\cite{Bayar:2016ftu,Guo:2015umn}. Indeed, one does observe an enhancement at the $KK\bar K$ threshold, as Fig.~\ref{fig:gammatilde} for $\tilde\Gamma(Kf_0^{K\bar K})$ shows. Of course, that amplitude contains not only the triangle, corresponding to the first loop of Fig.~\ref{fig:production} if formed with $Kf_0$ and kaon exchange. It also contains rescattering to all orders, and coupled channels. Such a full rescattering scheme, in which TSs are nothing but channel transitions with exchanged on-shell particles, has first been studied in Ref.~\cite{Sakthivasan:2024uwd}.
As the TS in the current case is situated just at threshold, we observe it as an enhanced cusp. Consequently, the on-shell condition is closest to being fulfilled for spectator momentum $p=0$, and, indeed, the amplitude is by far largest for small spectator momenta as Fig.~\ref{fig:gammatilde} demonstrates (red curves for the first channel, $Kf_0^{K\bar K}$). See also Fig.~\ref{fig:mov_cusp} in which this enhancement at $p=0$ becomes even more visible. The fact that the enhancement occurs for small spectator momenta $p$  for channels 1, 3, and 5 is also visible in the right column of Fig.~\ref{fig:gammatilde} in which the amplitude is plotted as a function of $p$. There, the green curves that peak around $p\approx 0$ correspond to intermediate energies at around $\sqrt{s}=1.48$~GeV where the threshold cusp is situated.

We have also conducted a numerical experiment by setting the tuning parameter from \cref{sec:isobars} to $x=1.5$. The $f_0$ resonance acquires then a width of over 100~MeV making the triangle condition meaningless. When plotting $\tilde \Gamma$ with this change, the cusp for channel 1, which is of size 1 as shown in the first panel of \cref{fig:gammatilde}, decreases to around 0.05; through coupled-channel effect the cusp of channel 5 ($\bar KK_2$) diminishes from 2 to 0.5 although the isobar in that channel is not altered, at all. This finding confirms the hypothesis that the $KK\bar K$ cusp is enhanced through the classic triangle effect, on one hand, and the complex $Kf_0$ threshold, on the other hand.

In summary, we have here three effects: 1) a dynamically generated pole which is, however, shielded by the cut of the complex $Kf_0$ threshold from the physical axis as discussed in \cref{sec:poletrajectory}; 2) a complex $Kf_0$ threshold that is also very close to the $KK\bar K$ threshold; 3)  a triangle singularity right at threshold. Of course, the $Kf_0$ threshold is just the other type of triangle singularity~\cite{Sakthivasan:2024uwd} produced by the Landau equations~\cite{Landau:1959fi}.

\section{Discussion and conclusions}
Three-body dynamics govern the properties of many strange mesons. We study here the $K(1460)$ resonance that can couple to three-body channels like $Kf_0$, with the narrow $f_0$ decaying to $\pi\pi$ or $K\bar K$.

The $Kf_0$ system can transition into itself or other channels like $Ka_0$ by exchanging an antikaon that can go on-shell right at the $KK\bar K$ threshold. This leads to the observation of a series of interesting phenomena: The strong attraction in the coupled $Kf_0$ and $Ka_0$ channels, enhanced through the near-on-shell exchanged kaon, leads to the formation of a dynamically generated state (``$K(1460)$'') emerging from the nonlinear meson-meson dynamics. This confirms previous work in the limit of mass-degenerate, stable isobars. While long-range forces from particle exchange are generally weak, the on-shell condition of the exchanged kaon does lead to a strong attraction in this case. 

The picture becomes more complicated once coupled-channel two-body amplitudes are used for the isobars, that represent available two-body scattering data (in a restriction to $S$-waves). The mass-degenerate $Kf_0$, $Ka_0$ thresholds move into the complex plane as $f_0$ and $a_0$ become resonances. On its way, the $Kf_0$ threshold not only drags the $K(1460)$ pole with it, but also shields it from the physical axis by its associated cut, making it much less visible. 

In the language of triangle singularities, the on-shell exchanged kaon causes a triangle singularity at the $KK\bar K$ threshold, coming from the almost ``on-shell'' $f_0$ resonance. In agreement with this picture, the enhancement of the amplitude should only occur for small outgoing spectator momenta, and it does, resulting in a very visible cusp. 

To connect these phenomena with observations, the formalism is embedded in an enlarged coupled-channel system constructed with manifest three-body unitarity, containing also the $\kappa$ and $\sigma$ resonances, as well as strangeness$-$2 kaon clusters. One of the channels is $K\pi\pi$ from the $\pi\pi$ decay of the $f_0$. We observe a pronounced enhancement in this channel, but also in the $K\pi\eta$ channel at around $\sqrt{s}\approx 1480$ MeV, for small spectator momenta. This originates from the previously discussed underlying enhancement at this energy in the $Kf_0^{K\bar K}$ channel, followed by a sub-channel $K\bar K\to\pi\pi$ transition (and analogously for the $K\pi\eta$ case). 
This enhancement at $\sqrt{s}\approx 1480$~MeV, observed in different final states, could be mistaken as a resonance signal, especially because the phase motions shown in Argand plots resemble that of a resonance. 

In future studies, the present amplitudes could be used to match to amplitudes extracted from experiment, or to fit experimental data directly --with additional steps like inclusion of higher-spin isobars. This would help to find out if the experimental signal related to the $K(1460)$ can be regarded as a resonant one, and if the latter corresponds to a hadronic molecule or genuine state.
In that case one would not necessarily expect the resonant shape to be concentrated at small spectator momenta as predicted here, but more evenly distributed. Alternatively, finding the ``resonance'' signal mostly at small spectator momenta would support the present hypothesis of it being mainly a consequence of a kinematic enhancement due to a triangle singularity and a complex threshold.


\bigskip

\begin{acknowledgments}
We thank Maxim Mai for useful discussions.
The work of MD is supported by the National Science Foundation (NSF) Grant No. 2310036. MD is also supported by the U.S. Department of Energy grant DE-SC0016582, and Office of Science, Office of Nuclear Physics under contract DE-AC05- 06OR23177.  
A.M.T, and K.P.K, gratefully acknowledge the partial support provided by the Brazilian agency CNPq (Conselho Nacional de Desenvolvimento Cient\'ifico e Tecnol\'ogico)(K.P.K: Grants No. 306461/2023-4; A.M.T: Grant No. 304510/2023-8). Further, A.M.T. thanks Fapesp (Funda\c c\~ao de Amparo \`a Pesquisa do Estado de S\~ao Paulo): Grant number 2023/01182-7.

\end{acknowledgments}
\footnotesize
\bibliography{BIB,NON-INSPIRE}
\clearpage
\appendix
\end{document}